\RequirePackage{fix-cm}
\documentclass{svjour3}            
\usepackage[round, sort]{natbib}
\setcitestyle{aysep={}}

\smartqed  
\usepackage{graphicx}
\usepackage[utf8]{inputenc}
\usepackage{amssymb}                  
\usepackage{amsmath}                  
\usepackage[squaren,Gray]{SIunits}    
\usepackage{numprint}                 
\usepackage{float}                    
\usepackage{subfigure}
\usepackage[bottom]{footmisc}
\usepackage{caption}
\usepackage{booktabs}                 
\DeclareUnicodeCharacter{2032}{}
\usepackage{color}
\usepackage{nomencl}
\makenomenclature

\usepackage{hyperref}

\usepackage[capitalise]{cleveref}

\journalname{Stochastic Environmental Research and Risk Assessment}

\begin{document}

\title{Comparison of Polynomial Chaos and Gaussian Process surrogates for  uncertainty quantification and correlation estimation of spatially distributed open-channel steady flows}
\titlerunning{Comparison of PC and GP surrogates for steady 1-D open-channel flows}        

\author{Pamphile T. Roy         \and
        Nabil El Mo\c{c}ayd	 \and
        Sophie Ricci		 \and
        Jean-Christophe Jouhaud \and
        Nicole Goutal         \and
        Matthias De Lozzo     \and
        Mélanie C. Rochoux
        }
\authorrunning{P.T.~Roy, N.~El Mo\c{c}ayd \emph{et al.}} 

\institute{P.T.~Roy \and J-C. Jouhaud\at
          CFD Team, CERFACS, 42 Avenue Gaspard Coriolis, 31057 Toulouse cedex 1, France \\
           \email{roy@cerfacs.fr}           
           \and
           N.~El Mo\c{c}ayd \and S.~Ricci \and M.~de Lozzo \and M.C.~Rochoux \at
           CECI, CERFACS -- CNRS, 42 Avenue Gaspard Coriolis, 31057 Toulouse cedex 1, France 
           \and
           N.~Goutal \at
           Laboratory for Hydraulics Saint-Venant (LHSV), EDF R\&D, 6 Quai Watier, 78401 Chatou, France
           }

\date{Received: 03/05/2017 / Accepted: 13/10/2017\\
This is a pre-print of an article published in Stochastic Environmental Research and Risk Assessment. The final authenticated version is available online at: https://doi.org/10.1007/s00477-017-1470-4}

\maketitle
\begin{abstract}
Data assimilation is widely used to improve flood forecasting capability, especially through parameter inference requiring statistical information on the uncertain input parameters (upstream discharge, friction coefficient) as well as on the variability of the water level and its sensitivity with respect to the inputs. For particle filter or ensemble Kalman filter, stochastically estimating probability density function and covariance matrices from a Monte Carlo random sampling requires a large ensemble of model evaluations, limiting their use in real-time application. To tackle this issue, fast surrogate models based on Polynomial Chaos and Gaussian Process can be used to represent the spatially distributed water level in place of solving the shallow water equations. This study investigates the use of these surrogates to estimate probability density functions and covariance matrices at a reduced computational cost and without the loss of accuracy, in the perspective of ensemble-based data assimilation. This study focuses on 1-D steady state flow simulated with MASCARET over the Garonne River (South-West France). Results show that both surrogates feature similar performance to the Monte-Carlo random sampling, but for a much smaller computational budget; a few MASCARET simulations (on the order of 10-100) are sufficient to accurately retrieve covariance matrices and probability density functions all along the river, even where the flow dynamic is more complex due to heterogeneous bathymetry. This paves the way for the design of surrogate strategies suitable for representing unsteady open-channel flows in data assimilation. \\

\end{abstract}

\begin{acknowledgements}
The financial support provided by CNES and EDF R\&D is greatly appreciated. The authors acknowledge Michael Baudin, Géraud Blatman, Anne Dutfoy, Bertrand Iooss and Anne-Laure Popelin from MRI (EDF R\&D) as well as Cédric Goeury from LNHE (EDF R\&D) for helpful discussions on uncertainty quantification and support on OpenTURNS.
\end{acknowledgements}

\newpage
\newcommand{\nomunit}[1]{
\renewcommand{\nomentryend}{\hspace*{\fill}
$\unit{}{#1}}$}

\nomenclature{$\boldsymbol{\zeta}$}{Standardized random variable}
\nomenclature{$\delta$}{Kronecker delta function}
\nomenclature{$\ell$}{Correlation length scale}
\nomenclature{$\gamma$}{Basis coefficient}
\nomenclature{$\pi$}{Correlation function}
\nomenclature{$\lambda$}{Singular value}
\nomenclature{$\mathbb{E}$}{Expectation operator}
\nomenclature{$\mathbb{V}$}{Variance}
\nomenclature{$\mathbf{\Lambda}$}{Rectangular singular value matrix}
\nomenclature{$\mathbf{C}$}{Snapshot covariance matrix}
\nomenclature{$\mathbf{h}$}{Output random vector of size $M$}
\nomenclature{$\mathbf{U}$}{Orthogonal square left singular matrix}
\nomenclature{$\mathbf{V}$}{Orthogonal square right singular matrix}
\nomenclature{$\mathbf{x}$}{Input random vector of size $d$}
\nomenclature{$\mathbf{x}^*$}{Input random vector not included in the training set $\mathcal{X}$}
\nomenclature{$\mathbf{Y}$}{Centred snapshot matrix}
\nomenclature{$\mathcal{D}_N$}{MASCARET simulation database of size $N$}
\nomenclature{$\mathcal{M}$}{MASCARET forward model operator}
\nomenclature{$\mathcal{N}$}{Normal distribution characterized by mean and STD}
\nomenclature{$\mathcal{U}$}{Uniform distribution characterized by minimum and maximum}
\nomenclature{$\mathbf{\Pi}$}{Correlation matrix evaluated for the database $\mathcal{D}_N$}
\nomenclature{$\mathcal{X}$}{Input training set of size $N \times d$}
\nomenclature{$\mathcal{Y}$}{Output training set of size $N \times M$}
\nomenclature{$\mu$}{Mean value}
\nomenclature{$\omega$}{Gaussian quadrature weight}
\nomenclature{$\Psi$}{Basis function}
\nomenclature{$\rho$}{Joint probability density function}
\nomenclature{$\sigma$}{STD}
\nomenclature{$\tau$}{Nugget effect}
\nomenclature{$\widehat{h}$}{Estimated water level}

\nomenclature{$a \in [a_{\text{in}}; a_{\text{out}}]$}{Curvilinear abscissa \nomunit{km}}
\nomenclature{$A$}{Hydraulic section \nomunit{m^2}}
\nomenclature{$c(\alpha)$}{$\alpha$-level tabulated value associated with $D$}
\nomenclature{$D$}{Kolmogorov-Smirnov statistic}
\nomenclature{$d$}{Uncertain space size}
\nomenclature{$h$}{Water level part of the hydraulic state $(h,Q)$ \nomunit{m}}
\nomenclature{$i$}{Surrogate decomposition index}
\nomenclature{$k$}{Snapshot index}
\nomenclature{$K_s$}{Strickler friction coefficient \nomunit{m^{1/3}\,s^{-1}}}
\nomenclature{$M = \numprint{14}$}{Number of observation stations}
\nomenclature{$N$}{Training set size}
\nomenclature{$N_{\text{ref}} = \numprint{100,000}$}{Validation set size}
\nomenclature{$P$}{Wet perimeter \nomunit{m}}
\nomenclature{$Q$}{Discharge part of the hydraulic state $(h,Q)$ \nomunit{m^{3}\,s^{-1}}}
\nomenclature{$Q_2$}{Predictive squared correlation coefficient}
\nomenclature{$r$}{Number of terms in surrogate model}
\nomenclature{$R$}{Hydraulic radius \nomunit{m}}
\nomenclature{$RC$}{Upstream water level-discharge local rating curve}
\nomenclature{$S_0$}{Channel slope \nomunit{m\,km^{-1}}}
\nomenclature{$S_f$}{Friction slope}
\nomenclature{$S_i$}{First order Sobol' index}
\nomenclature{$S_T$}{Total order Sobol' index}
\nomenclature{$W$}{River width \nomunit{m}}

\nomenclature{CDF}{Cumulative Distribution Function}
\nomenclature{DA}{Data Assimilation}
\nomenclature{EnKF}{Ensemble Kalman Filter}
\nomenclature{GP}{Gaussian Process}
\nomenclature{MC}{Monte Carlo}
\nomenclature{PC}{Polynomial Chaos}
\nomenclature{PDF}{Probability Density Function}
\nomenclature{PGP}{POD-based Gaussian Process}
\nomenclature{POD}{Proper Orthogonal Decomposition}
\nomenclature{PF}{Particle Filter}
\nomenclature{RMSE}{Root Mean Square Error}
\nomenclature{SA}{Sensitivity Analysis}
\nomenclature{STD}{STandard Deviation}
\nomenclature{SVD}{Singular Value Decomposition}
\nomenclature{SWE}{Shallow Water Equation}
\nomenclature{UQ}{Uncertainty Quantification}

\printnomenclature

\newpage
\section{Introduction}\label{sec:intro}

The predictive skills of hydraulic models have greatly increased with advances in free surface flow numerical modelling and computational resources. Real-time flood forecasting relies on the use of sparse in situ observations as well as imperfect hydrology or hydraulic models usually solving the 1-D Shallow Water Equations (SWE). Assessing the predictive capabilities of these hyperbolic partial differential equations remains an important challenge as public safety and water resource management are at stake~\citep{weerts2011}. SWE solve for spatially varying water level and river discharge (referred to as the river state) using physical parameters (e.g.~friction coefficients), initial conditions and boundary conditions described as a hydrograph or limnigraph. These input data are subject to epistemic uncertainties due to an imperfect knowledge of the river properties as well as to aleatory uncertainties related to environmental and meteorological intrinsic hazards. Both types of errors translate into uncertainties in the simulated river state, thus preventing the hydraulic model from being effective in forecast mode. In practice, these uncertainties can be reduced when complementary data become available.

Data assimilation (DA) offers a convenient framework to reduce model uncertainties by combining observations with the model simulation taking into account errors in both sources of information. Prior to DA, the main sources of uncertainties should be identified and included in the control vector; this is achieved with a sensitivity analysis (SA) study that allows to classify uncertainties in the inputs with respect to their impact on the model outputs, for instance in terms of variance using Sobol' indices~\citep{iooss2016}. Several studies in the framework of hydraulics demonstrated the merits of DA~\citep{barthelemy2015,cloke2009,dechant2011,habert2016,moradkhani2005} to provide a more accurate river state.
This is achieved by inferring an optimal set of parameters (e.g.~river and floodplain friction coefficients, upstream and lateral river discharge, bathymetry) and/or by simulating a more accurate river state, thus constituting a paradigm shift for real-time flood forecasting. 
Ensemble-based methods such as the Ensemble Kalman Filter (EnKF)~\citep{durand2008,moradkhani2005} and the Particle Filter (PF)~\citep{matgen2010,parrish2012} are popular algorithms; they articulate as a two-step procedure derived from Bayesian inference: (1)~a forecast step to sample the uncertain inputs and propagate the uncertainty through the model, thus providing an ensemble of river states; and (2)~an analysis step to weight each member or particle of the ensemble based on its discrepancies to the available observations and to derive in the case of parameter estimation, a correction on the inputs that is then propagated to the river states by model integration. In the EnKF algorithm, the weights are provided by the stochastic estimation of covariance matrices between errors in model inputs and outputs. In contrast, in the PF algorithm, the weights correspond to likelihoods associated with the probability density function (PDF) of the control vector conditioned upon the observations; PF provides an alternative to EnKF when the model is subject to strong non-linearity and non-Gaussian errors.  

Monte Carlo (MC) techniques are generic, robust and easily portable on massively parallel supercomputers; yet they remain computationally expensive due to their slow convergence rate scaling as the inverse of the square root of the number of particles~\citep{lixiu2008}. As shown in~\citet{barthelemy2015} and \citet{bozzi2015}, a large number of forward model evaluations should be carried out to converge the stochastic evaluation of error statistics such as PDFs, Sobol' indices and covariance matrices. There is therefore a need to develop efficient and robust uncertainty quantification (UQ) methods in the context of DA for hydraulics to limit (1)~the number of significant sources of uncertainties and (2)~the computational cost of quantifying uncertainties on the river state, e.g.~moments (mean, covariance) and PDF, while preserving the accuracy of the mapping $\mathcal{M}$ between the uncertain inputs $\mathbf{x}$ and the vector of $M$ river water heights $\mathbf{h}$:
\begin{equation}
\mathbf{x} \in \mathbb{R}^d \quad \rightarrow \quad \mathbf{h} = \mathcal{M}(\mathbf{x})\in\mathbb{R}^M.
\end{equation}
The key idea of non-intrusive UQ methods is to build a cost-effective surrogate, also called reduced model, metamodel or response surface, mimicking the mapping $\mathcal{M}$ to perform UQ and SA steps~\citep{iooss2010,iooss2016,lamboni2011,lemaitreknio2010,saltelli2007,storlie2009}. Formulating the surrogate model relies on a limited number of forward model integrations referred to as the training sample $(\mathcal{X}, \mathcal{Y})=\left(\mathbf{x}^{(k)},\mathbf{h}^{(k)}\right)_{1\leq k\leq N}$ where $N$ is the training sample size and $\mathbf{h}^{(k)} := \mathcal{M}(\mathbf{x}^{(k)})$ corresponds to the deterministic integration of the forward model $\mathcal{M}$ as a black box for the $k$th set of input parameters $\mathbf{x}^{(k)}$. Several surrogate models are found in the literature, among whom generalized linear models, polynomial models, splines, polynomial chaos expansions, artificial neural networks, Gaussian process models, ...

Polynomial chaos (PC) approach has received much attention lately~\citep{dubreuil2014,sudret2008,xiu2010,xiu2002}. The PC surrogate model is formulated as a polynomial expansion, in which the basis is defined according to the distribution of the uncertain inputs $\mathbf{x}$ and the coefficients relate to the statistics of the output $\mathbf{h}$. The coefficients are computed so as to fit the training set $(\mathcal{X}, \mathcal{Y})$, either using regression or spectral projection methods. The merits of PC surrogates were demonstrated in various fields, e.g.~structural mechanics~\citep{dubreuil2014,berveiller2005}, computational fluid dynamics~\citep{hosder2006,lucor2007,saad2007phd}, hydrology~\citep{deman2015}, hydraulics~\citep{ge2008}, wildfires~\citep{rochoux2014}. A complementary approach between PC and EnKF was presented in~\citet{lixiu2009} and tested in the framework of wildfire behaviour forecasting in~\citet{rochoux2014}. A PC surrogate was used in place of the forward model to estimate the Kalman gain and thereby design a cost-effective EnKF inferring new estimates of the input parameters (e.g.~biomass moisture and fuel aspect ratio) reducing bias in the fire front location prediction. The resulting algorithm achieved convergence in spite of the non-linear surface response, showing promising results for environmental risk monitoring. 

Gaussian processes (GP) that are strongly related to Kriging in geostatistics also are of increasing interest~\citep{legratiet2017,lockwood2012,rasmussen2006}. The GP formalism treats the forward model response as a realization of a Gaussian stochastic process indexed by $\mathbf{x}$ and fully characterized with mean and covariance functions conditioned by the training set $(\mathcal{X}, \mathcal{Y})$. The GP surrogate is built first, by defining a covariance kernel function between output values and a trend function and then, by estimating the hyperparameters (e.g.~variance, characteristic length scale) that provide a good fit to the training set. GP surrogates were introduced in the context of SA for estimating Sobol' indices~\citep{oakley2004,marrel2009,legratiet2014}.

PC and GP surrogate models have recently been compared for UQ and SA studies~\citep{legratiet2017,owen2015,schoebi2015}. \citet{owen2015} evaluated the performance of each type of surrogate in terms of output mean, variance and PDF estimation. \citet{legratiet2017} compared Sobol' indices with applications in structural mechanics; PC and GP surrogate models were found to feature a similar accuracy with respect to the $Q_2$ validation coefficient for a given size of the training set. They also emphasized that the ranking between PC and GP approaches remains problem-dependent. Since the application of PC surrogate models to hyperbolic conservation laws such as SWE remains a challenging task~\citep{despres2013,birolleau2014}, it is of great interest to evaluate the performance of PC and GP approaches when applied to open-channel flow simulation in fluvial or critical configurations.    

In this paper, our objective is to evaluate the performance of PC and GP surrogates mimicking the behaviour of 1-D open-channel flows for UQ and SA steps that are important in the design of EnKF and PF algorithms. The quantity of interest is the 1-D water level discretized along 50~km of the Garonne River (South-West France). As a preliminary step, the focus is made on steady flows and the flow remains fluvial. The treatment of strong non-linearity induces by fluvial/critical transitions is beyond the scope of this study. The main sources of uncertainties are the upstream discharge that is constant in steady state conditions and the friction coefficient that is a piecewise constant function. Under these hypotheses, the present work features scalar inputs and spatially varying outputs due to the heterogeneous bathymetry of the Garonne River. A convergence study is first carried out to determine the size of the training set $N$ that is required to build a valid surrogate using either PC expansion or GP model. GP and PC surrogates are then compared when a computational budget is set, i.e.~for the same number of snapshots (also called simulations in the paper) $N$ used to construct each surrogate. The comparison is carried out with respect to the following metrics on $\mathbf{h}$: PDF that is of interest for Bayesian inference and PF algorithms; spatially varying Sobol' indices (associated with the correlation between each uncertain input and the spatially distributed output) and correlation matrix (associated with the spatial correlation of the output) that are of interest for variational and EnKF algorithms.

The structure of the paper is as follows. Section~\ref{sec:hydrau} introduces the 1-D SWE solved using MASCARET and the Garonne River case study. The PC and GP techniques are detailed in Sect.~\ref{sec:UQmethod} along with error metrics. Section~\ref{sec:UQresults} presents the results of the comparative study between PC and GP with respect to a MC reference. Conclusions and perspectives are given in Sect.~\ref{sec:ccl}.

\section{Hydraulic Modelling}\label{sec:hydrau}

The SWE represent the dynamics of open-channel flows, typically in rivers with small bathymetry variations~\citep{horritt2002}. They form a hyperbolic system of partial differential equations that characterize subcritical and supercritical flows subject to hydraulic jumps. Here, we only deal with subcritical flows, in the plain in the downstream part of the Garonne River.

\subsection{1-D shallow water equations (SWE)}

We consider a 1-D hydraulic model commonly used in hydraulic engineering and flood forecasting. The stream channel is described by a hydraulic axis corresponding to the main flow direction, implying that the river channel is represented by a series of cross-sections (or profiles) identified by a curvilinear abscissa $a$ ranging from $a_{\text{in}}$ upstream of the river to $a_{out}$ downstream. 1-D SWE are derived from mass conservation and momentum conservation. The equations are written in terms of discharge (or flow rate) $Q$~(m$^3$\,s$^{-1}$) and hydraulic section $A$~(m$^2$) that relates to water level (or water height) $h$~(m) such that $A \equiv A(h)$. The non-conservative form of the 1-D SWE for non-stationary flow reads~\citep{thual2010} 
\begin{eqnarray}
\left\{
\begin{array}{l}
\displaystyle{\partial_t A(h) + \partial_a Q=0} \\
\displaystyle{\partial_t Q +  \partial_a\left(\frac{Q(h)^2}{A(h)}\right) + g\,A(h)\,\partial_a h - g\,A(h)\,(S_0-S_f) = 0} \\
\end{array}
\right.
\label{eq:SWE}
\end{eqnarray}
with $g$ the gravity, $S_0$ the channel slope and $S_f$ the friction slope. In the present study, the SWE are combined with the Manning-Strickler formula to parameterize the friction slope $S_f$ such as:
\begin{equation}
S_f = \frac{Q^2}{K_s^2\,A(h)^{2}\,R(h)^{4/3}},   
\label{eq:friction}
\end{equation}
where $R(h) =A(h)/P(h)$~(m) is the hydraulic radius written as a function of the wet perimeter $P(h)$, and where $K_s$~(m$^{1/3}$\,s$^{-1})$ is the Strickler friction coefficient. The pair $(h,Q)$ forms the hydraulic state varying in time and space. For steady flows, Eq.~\ref{eq:SWE} simplifies to: 
\begin{eqnarray}
\left\{
\begin{array}{l}
\displaystyle{\partial_a Q=0} \\ \\
\displaystyle{\partial_a h =\frac{(S_0-S_f)}{1-Fr^ 2}}\\
\end{array}
\right.
\label{eq:backwater}
\end{eqnarray}
where $Fr$ is the dimensionless Froude number 
\begin{equation}
Fr^2 = \frac{Q^2}{gA^3} \frac{\partial{A}}{\partial{h}}.
\label{eq:froude}
\end{equation}

The smooth solutions for Eq.~\ref{eq:backwater} are called {\it backwater curves} when the downstream boundary condition is prescribed in a deterministic way. To solve Eq.~\eqref{eq:backwater}, the following input data are required: bathymetry, upstream and downstream boundary conditions, lateral inflows, roughness coefficients and initial condition for the hydraulic state. The imperfect description of this data translates into errors in the simulated hydraulic state $(h,Q)$. To understand the structure of these errors, it is of prime importance to determine which input variables contribute, and to what extent, to the variability in the hydraulic state at different curvilinear abscissas $a$ along the river channel, for instance via a SA study (Sect.~\ref{sec:SA}).

We use the MASCARET software to simulate the 1-D SWE in Eq.~\eqref{eq:SWE} and predict $(h, Q)$ along the discretized curvilinear abscissa of the hydraulic network $a \in [a_{\text{in}}, a_{\text{out}}]$. The SWE are solved here with the steady kernel of MASCARET based on a finite difference scheme~\citep{goutal2002,goutal2012}, meaning that $(h,Q)$ only varies in space. MASCARET is part of the TELEMAC-MASCARET open-source modelling package developed at EDF (\textit{Électricité de France} R\&D) in collaboration with CEREMA (\textit{Centre d'Étude et d'expertise sur les Risques, l'Environnement, la Mobilité et l'Aménagement}); it is commonly used for dam-break wave simulation, reservoirs flushing and flooding. 

\subsection{Garonne River Case Study}

\begin{figure*}[htb]
\centering
\subfigure[]{\includegraphics[scale=0.35]{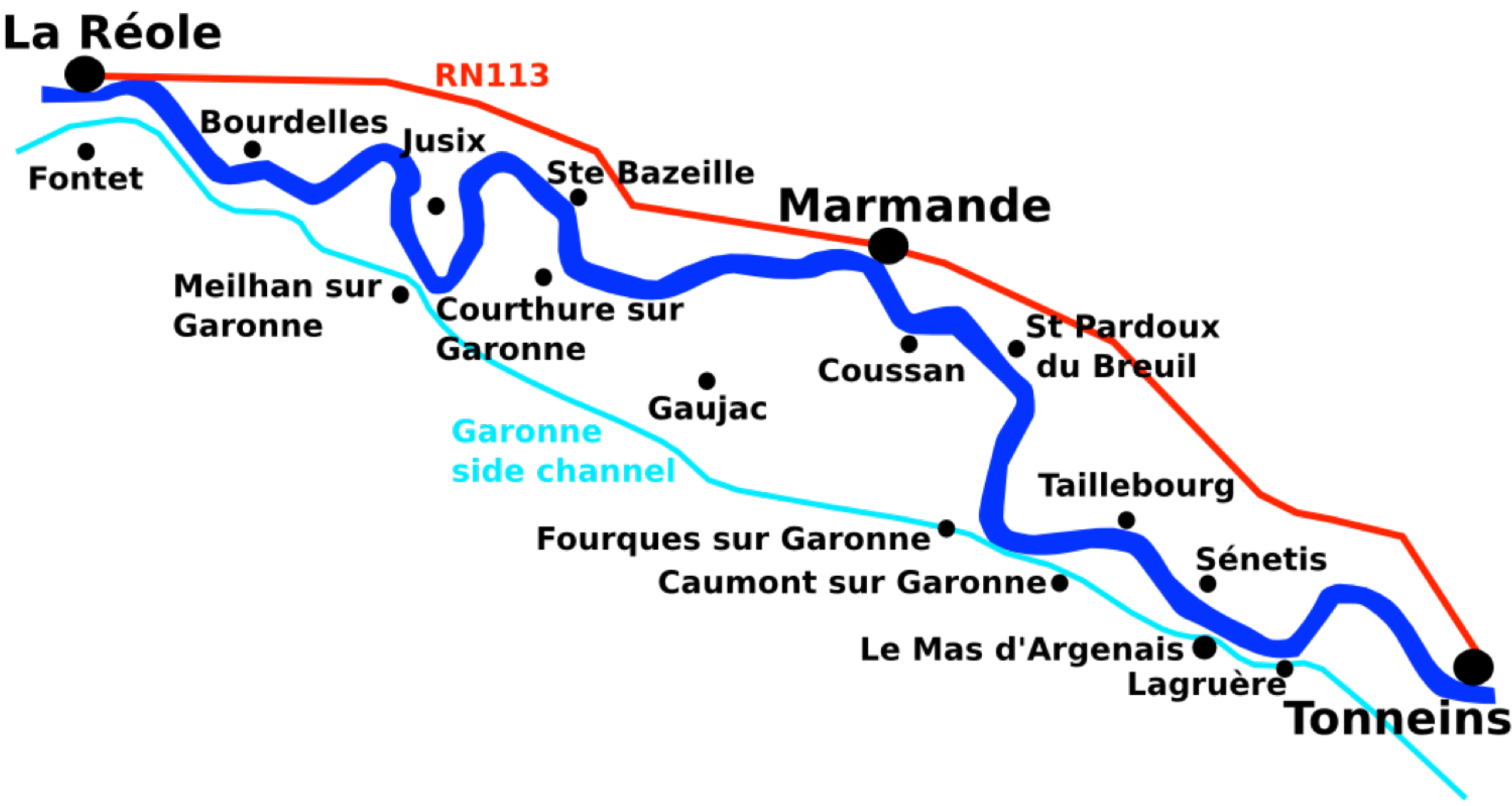}}
\subfigure[]{\includegraphics[scale=0.55]{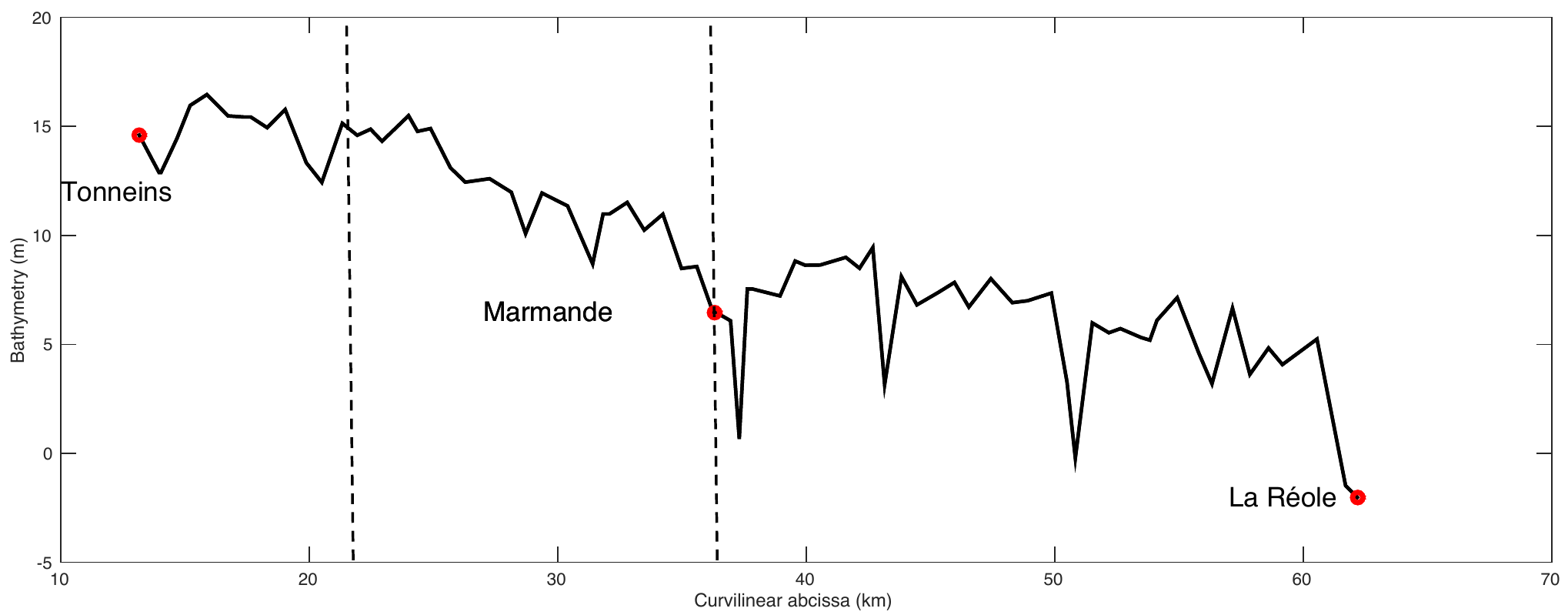}}
\caption{Garonne River case study (South-West France). (a)~Reach between Tonneins (upstream, $a_{\text{in}} = 13$~km) and La Réole (downstream, $a_{\text{out}} = 62$~km) with Marmande located at $a = 36$~km. (b)~Bathymetry profile along the curvilinear abscissa $a$~(km) between Tonneins and La Réole. The Strickler friction coefficient $K_s$ spatially varies as a constant piecewise function; the changes in the value of $K_s$ are indicated by vertical dashed lines.}
\label{fig:garonne}
\end{figure*}

The present study is carried out on a real hydraulic network over the Garonne River in South-West France. The Garonne river flows along 647~km from the Pyrenees to the Atlantic Ocean draining an area of $55,000$~km$^2$ (corresponding to the fourth-largest river in France). The present study focuses on a 50 km reach from Tonneins ($a_{\text{in}} = 13$~km) to La Réole ($a_{\text{out}} = 62$~km) with one observing station at Marmande ($a = 36$~km) as presented in Fig.~\ref{fig:garonne}a. The mean slope over the reach is $S_0 = 3.3$~m\,km$^{-1}$; the mean width of the river is $W = 250$~m; the bank-full discharge is approximately $\overline{Q} = 2550$~m$^3$\,s$^{-1}$. Despite the existence of active floodplains, this reach can be accurately modelled by a 1-D hydraulic model using Eq.~\eqref{eq:SWE}~\citep{besnard2011}. 

Figure~\ref{fig:garonne}b presents the non-uniform bathymetry profile along the 50 km reach, interpolated from $83$ on-site bathymetry cross-sections. Friction for the river channel and its floodplain is prescribed over three zones separated by dashed lines. The Strickler coefficients $K_{s_1}$, $K_{s_2}$ and $K_{s_3}$ are used to characterize friction through Eq.~\eqref{eq:friction} and are uniform per zone. The observing station at Marmande is located at the beginning of the third zone associated with $K_{s_3}$. 

The upstream steady boundary condition is prescribed by $Q(a_{\text{in}}) = Q_{\text{in}}$; the discharge $Q$ is constant along the reach ($Q = Q_{\text{in}}$). The downstream boundary condition is prescribed with a local rating curve $RC$ established at La R\'eole that sets $h(a_{\text{out}}) = RC(Q_{\text{out}}) = h_{\text{out}}$. The hydraulic model has been calibrated using channel and floodplain roughness coefficients as free parameters~\citep{besnard2011}. 

\subsection{Sources of uncertainties and Quantity of Interest}\label{sec:database}

UQ and DA for flood forecasting are essential to ensure the predictive capability of the surrogate model at the observing stations such as Marmande. Previous work~\citep{elmocaydEMA} has shown that in steady state conditions and given some assumptions on the statistics of the input uncertain variables, the sensitivity of the hydraulic state at Marmande to $K_{s_3}$ is predominant; the sensitivity at Marmande to $K_{s_1}$ and $K_{s_2}$ is negligible. Hence, the main sources of uncertainties taken into account here are the upstream mass flow rate $Q$ and the Strickler coefficient $K_{s_3}$. We denote by $\mathbf{x} = (Q, K_{s_3})$ the random vector of size $d = 2$. From expert knowledge, $Q$ and $K_{s_3}$ are considered as independent random variables where $Q$~(m$^3$\,s$^{-1}$) follows the normal distribution $\mathcal{N}(4031,400)$ and $K_{s_3}$~(m$^{1/3}$\,s$^{-1}$) follows the uniform distribution $\mathcal{U}(15,60)$. 

MASCARET provides as output the water height over 463 cross-sections for the Garonne case. In this work, we focus on the water height at $M = 14$ stations evenly distributed along the 50 km reach, among which Marmande at $a = 36$~km. We denote by $\mathbf{h}$ the vector of $M$ observed water levels for one realization of MASCARET. Note that the UQ methods we present in Sec.~\ref{sec:UQmethod} are highlighted in this work for $M = 14$ but they can easily be applied to a case where the size of the state vector $M$ is much larger than the size of the training set $N$.

A database noted $\mathcal{D}_{N_{\text{ref}}}$ and containing $N_{\text{ref}} = \numprint{100,000}$ MASCARET simulations was compiled as a reference for the study. Each simulation corresponds to a different pair of inputs ($Q, K_{s_3}$) resulting from a MC random sampling following $Q\sim\mathcal{N}(4031,400)$ and $K_{s_3} \sim \mathcal{U}(15,60)$. In the present study $D_{N_{\text{ref}}}$ is partly used to build the PC and GP surrogates via the training set 
$\mathcal{D}_N = (\mathcal{X}$, $\mathcal{Y})\subset \mathcal{D}_{N_{\text{ref}}}$ of size $N$; it is fully used to validate them a posteriori over a large ensemble (Sec.~\ref{sec:error}).

\section{Surrogate Models}\label{sec:UQmethod}

Two types of surrogate models are used in the present work to approximate the behaviour of MASCARET with respect to $\mathbf{x} = (Q, K_{s_3})$ (the size of the input uncertainty space is $d = 2$): PC expansion on the one hand, GP regression combined with Proper Orthogonal Decomposition (POD) further denoted by pGP on the other hand. The common idea of PC and pGP approaches is to design for each curvilinear abscissa $a \in \{a_1, \cdots, a_M\}$ a surrogate water level with a weighted finite sum of basis functions:
\begin{equation}
\widehat{h}_a\left(\mathbf{x}\right) = \displaystyle\sum_{i = 0}^{r}\,\gamma_{a,i}\,\Psi_{i}\left(\mathbf{x}\right),
\label{eq:SurrogateForm}
\end{equation}
where the coefficients $\gamma_{a,i}$ and the basis functions $\Psi_i$ are calibrated by the training set $\mathcal{D}_N$. For PC expansion, the maximum order of the decomposition and the basis functions are first chosen, then a spectral projection strategy is implemented to compute the decomposition coefficients in the polynomial basis. The pGP strategy is two-fold: the first step consists in transforming the sampled output space to an orthogonal one, eventually reducing its dimension; the second step consists in interpolating with a GP approach the principal components of this new space to express water level in the basis for any friction and discharge values. It should be noted that the GP model could be replaced by any other interpolator. The main difference between PC (Sect.~\ref{sec:PC}) and pGP (Sect.~\ref{sec:GP}) approaches stands in the nature of these models: pGP interpolates the training points and captures local variations, while PC is a regression method focusing on the global behaviour of the model. Basis functions and calibration methods also differ between these two approaches.

\subsection{Polynomial Chaos (PC) surrogate model}\label{sec:PC}

The algorithm to build the PC surrogate proceeds as follows:
\begin{enumerate}
\item choose the polynomial basis $\lbrace\Psi_{i}\rbrace_{i\geq 0}$ according to the assumed PDF of the inputs $\mathbf{x} = (Q, K_{s_3})$,
\item choose the total polynomial degree $P$ according to the complexity of the physical processes,
\item truncate the expansion to $r_{\text{pc}}$ terms to keep the predominant information given by the forward model using standard truncation strategy ($r_{\text{pc}}$ depends on $d$ and $P$),
\item apply spectral projection strategy (i.e.~Gaussian quadrature rule) to compute the coefficients $\lbrace\gamma_{a,i}\rbrace_{i\in\mathbb{N}^d\atop|i|\leq P}$ for each curvilinear abscissa $a$ using $N = (P+1)^{d}$ snapshots from the simulation database $\mathcal{D}_{N_{\text{ref}}}$,
\item formulate the surrogate model $\mathcal{M}_{\text{pc}}$ at each curvilinear abscissa $a$, which can be evaluated for any new pair of parameters $\mathbf{x}^* = (Q^*, K_{s_3}^*)$.
\end{enumerate}
Note that we use standard truncation and projection strategies presented in~\cite{lemaitreknio2010} and \cite{xiu2010}.

\subsubsection{Polynomial basis}

Each component of the random vector $\mathbf{x}$ defined in the input physical space is standardized and noted $\boldsymbol{\zeta}$ in the following way: $\zeta_i=\frac{x_i-\mu_i}{\sigma_i}$ where $\mu_i=N^{-1}\sum_{k=1}^Nx_i^{(k)}$ and $\sigma_i=\sqrt{(N-1)^{-1}\sum_{k=1}^N\left(x_i^{(k)}-\mu_i\right)^2}$. Assuming that the SWE solution is of finite variance, each component $h_a$ of the water level vector $\mathbf{h}$ can be considered as a random variable for which there exists a polynomial expansion of the form~\eqref{eq:SurrogateForm} that represents how the water level $h_a$ varies according to changes in $Q$ and $K_{s_3}$. 

$h_a$ is projected onto a stochastic space spanned by the orthonormal polynomial functions $\lbrace\Psi_{i}\rbrace_{i\geq 0}$. These functions are orthonormal with respect to the joint density $\rho(\boldsymbol{\zeta})$, i.e.
\begin{equation}
\int_{Z}\,\Psi_i(\boldsymbol{\zeta})\,\Psi_j(\boldsymbol{\zeta})\,\rho(\boldsymbol{\zeta})\,\mathrm{d}\boldsymbol{\zeta} = \delta_{ij},
\label{eq:pc_innerproduct}
\end{equation}
with $\delta_{ij}$ the Kronecker delta function and $Z \subseteq \mathbb{R}^d$ the space in which $\boldsymbol{\zeta}$ evolves. In practice, the orthonormal basis is built using the tensor product of 1-D polynomial functions: $\Psi_i=\Psi_{i,1}\ldots\Psi_{i,d}$ where $i$ is the multi-index $(i_1,\ldots,i_d)\in\{0,1,\cdots,P\}^d$. The choice for the basis functions depends on the probability measure of the random variables. According to Askey's scheme, the Hermite polynomials form the optimal basis for random variables following the standard Gaussian distribution, and the Legendre polynomials are the counterpart for the standard uniform distribution~\citep{xiu2002}. 

\subsubsection{Truncation Strategy}

In practice, the sum in Eq.~\eqref{eq:SurrogateForm} is truncated to a finite number of terms $r_{\text{pc}}$. Using a standard truncation strategy $r_{\text{pc}}$ is constrained by the number of random variables $d$ and by the total polynomial degree $P$ as:
\begin{equation}
r_{\text{pc}} = \frac{(d + P)!}{d!\,P!},
\label{eq:pc_order}
\end{equation}
meaning that all polynomials involving the $d$ random variables of total degree less or equal to $P$ are retained in the PC expansion. The PC approximated water level at curvilinear abscissa $h_{\text{pc}}(a)$ is formulated as:
\begin{equation}
\widehat{h}_{\text{pc},a}(\mathbf{x}) := \mathcal{M}_{\text{pc},a}(\boldsymbol{\zeta}) = \displaystyle\sum_{i\in\mathbb{N}^d\atop|i|\leq P}\,\gamma_{a,i}\,\Psi_i\left(\boldsymbol{\zeta}\right).
\label{eq:SurrogatePC}
\end{equation}
Note that for small $d$, advanced truncation strategies that consist in eliminating high-order interaction terms or using sparse structure~\citep{blatman2009phd,migliorati2013} are not necessary.

\subsubsection{Spectral projection strategy}

We focus here on non-intrusive approaches to numerically compute the coefficients $\lbrace\gamma_{a,i}\rbrace_{i\in\mathbb{N}^d\atop |i|<P}$  in Eq.~\eqref{eq:SurrogatePC} using $N$ snapshots from $\mathcal{D}_N$. The spectral projection relies on the orthonormality property of the polynomial basis. The $i$th coefficient $\gamma_{a,i}$ is computed using Gaussian quadrature as:
\begin{equation}
\gamma_{a,i} = <h_a,\Psi_i> \,\cong\,\displaystyle\sum_{k = 1}^{N}\,h_a^{(k)}\,\Psi_i(\boldsymbol{\zeta}^{(k)})\,w^{(k)},
\label{eq:pc_quadrature}
\end{equation}
where $\mathbf{h}^{(k)} = \mathcal{M}(\mathbf{x}^{(k)})$ is the snapshot $\mathcal{D}_N$ corresponding to the MASCARET simulation for the $k$th quadrature root $\mathbf{x}^{(k)}$ of $\Psi_i$ (in the physical space), and where $w^{k}$ is the weight associated with $\mathbf{x}^{(k)}$. $(P+1)$ is the number of quadrature roots required in each uncertain direction to ensure an accurate calculation of the integral $<h_a,\Psi_i>$. Hence, $N = (P+1)^2$ for the PC surrogates built in this study.

\subsection{POD-based Gaussian Process (pGP) surrogate}\label{sec:GP}

The algorithm to build a Gaussian Process surrogate relies on a POD and proceeds as follows:
\begin{enumerate}
\item choose the size of the training set $N$,
\item draw $N$ samples (or snapshots) in the input random space $\mathbf{x} = (Q, K_{s_3})$ with Halton's low discrepancy sequence{\color[rgb]{0.9,0.2,0.16}\footnote{From a Halton sequence, the nearest points (standardized Euclidean distance) of the database $\mathcal{D}_N$ are chosen.}} from the simulation database $\mathcal{D}_N$,
\item formulate the centred snapshot matrix $\mathbf{Y}$ from the $N$ water level snapshots,
\item achieve a POD on $\mathbf{Y}$ using the snapshot method to derive the basis vectors $\lbrace\Psi_{i}\rbrace$ and the corresponding coefficients $\lbrace\gamma_{a,i}\rbrace$ (any snapshot can be expressed as a linear combination of the basis vectors and coefficients),
\item replace each basis vector $\lbrace\Psi_i\rbrace$ associated to the $N$ snapshots by a basis function via Gaussian Process regression for any $\mathbf{x}^*$, 
\item formulate the surrogate model $\mathcal{M}_{\text{gp}}$ for water level at each curvilinear abscissa $a$, which can be evaluated for any $\mathbf{x}^*$.
\end{enumerate}
Note that we follow the choices made by~\cite{braconnier2011}. 

\subsubsection{Snapshot method}\label{sec:POD}

The key idea of the snapshot method~\citep{sirovich1987} is to achieve a POD of the centred snapshot matrix $\mathbf{Y} \in \mathbb{M}_{M,N}(\mathbb{R})$, which gathers the water level computed at each curvilinear abscissa for the $N$ snapshots, from which the sample mean is subtracted. For simplicity purpose, the water level anomaly at curvilinear abscissa $a$ is denoted $h$ in the following. Thus, $\mathbf{Y} = \left(h_{a_i}^{(j)}\right)_{1\leq i \leq M \atop 1\leq j \leq N}$. The snapshots correspond to the column vectors; the $k$th snapshot of size $M = 14$ is denoted by $\mathbf{h}^{(k)}$.

Based on many observations of a random vector, the POD gives the orthogonal directions of largest variances (or modes) in the probabilistic vector space in order to reduce the vector space dimension~\citep{chatterjee2000}. Note that for simplicity purpose, the adjective {\it centred} is dropped in the following when referring to the centred snapshot matrix $\mathbf{Y}$.

The POD of the snapshot covariance matrix $\mathbf{C} = N^{-1}\,\mathbf{Y}^{\mathrm{T}}\,\mathbf{Y}\in \mathbb{M}_N(\mathbb{R})$ is equivalent to the Singular Value Decomposition (SVD) of the snapshot matrix $\mathbf{Y}$:
\begin{equation}
\mathbf{Y} = \mathbf{U}\,\mathbf{\Lambda}\,\mathbf{V}^{\mathrm{T}} = \displaystyle\sum_{k = 1}^{r_{p}}\,\lambda_k\,\mathbf{u}_k\,\mathbf{v}_k^{\mathrm{T}},
\end{equation} 
where $\mathbf{U} \in \mathbb{M}_M(\mathbb{R})$ is an orthogonal matrix diagonalizing $\mathbf{Y}\mathbf{Y}^{\mathrm{T}}$ ($\mathbf{u}_k$, the $k$th column of $\mathbf{U}$, is a left singular vector of $\mathbf{Y}$), where $\mathbf{V} \in \mathbb{M}_N(\mathbb{R})$ is an orthogonal matrix diagonalizing $\mathbf{Y}^{\mathrm{T}}\mathbf{Y}$ ($\mathbf{v}_k$, the $k$th column of $\mathbf{V}$, is a right singular vector of $\mathbf{Y}$), and where $\mathbf{\Lambda} \in \mathbb{M}_{M,N}(\mathbb{R})$ is a rectangular diagonal matrix including $r_{p}=\min(M,N)$ singular values on its diagonal. The singular values $\lbrace \lambda_k \rbrace_{1 \leq k \leq r_{p}}$ are the square roots of the eigenvalues of $\mathbf{C}$. 
Note that in this study, we do not reduce further the rank of the snapshot matrix $\mathbf{Y}$. Since the number of stations ($M = 14$) is lower than the size of the training set $N$, the rank of $\mathbf{Y}$ is here $r_{p} = M$.

For a given curvilinear abscissa $a$, any snapshot $h_a(\mathbf{x}^{(k)})$ can then be retrieved as a linear combination of $r_{p}$ modes $\{\Psi_i\}_{1\leq i \leq r_p}$:
\begin{equation}\label{eq:gppod}
h_a(\mathbf{x}^{(k)}) 
= (\mathbf{U}\,\mathbf{\Lambda}\,\mathbf{V}^{\mathrm{T}})_{ak} 
= U_{a:}(\mathbf{\Lambda}\,\mathbf{V}^T)_{:k}=\sum_{i=1}^{r_{p}}\,\gamma_{a,i}\,\Psi_i(\mathbf{x}^{(k)}),
\end{equation}
where for any $i\in\{1,\ldots,M\}$, $\gamma_{a,i}:=U_{a,i}$ and $\Psi_i(\mathbf{x}^{(k)}):=(\boldsymbol{\Lambda}\mathbf{V}^T)_{i,k}$. 

In the present study, the size of the state vector $M$ is smaller than the size of the training set $N$, implying that there are $r_p = M$ non-zero singular values. This is usually not the case, but this does not change the methodology. As the objective is to compare PC and pGP and as there is a limited number of stations $M$, all modes are kept to avoid the loss of information, there is no need to reduce the rank of the singular matrix. Note that when moving to 2-D or 3-D cases (beyond the scope of the present work), the size of the state vector could be of the order of thousands components, making dimension reduction necessary. 

\subsubsection{Regression Procedure}\label{sec:regression}

Our objective is now to generalize the data set $\lbrace\Psi_i(\mathbf{x}^{(k)})\rbrace_{1\leq k \leq N}$ associated with the design of experiments $\mathcal{X}$ to any new input vector $\mathbf{x}^*$, in order to approximate the model output $h_a$ at any curvilinear abscissa $a$. We thus propose the following surrogate model based on GP regression and Eq.~\eqref{eq:gppod}:
\begin{equation}\label{pgpak}
\widehat{h}_{\text{pgp},a}(\mathbf{x}^*) = \sum_{i=1}^{r_{p}}\gamma_{a,i}\,\Psi_{\text{gp},i}(\mathbf{x}^*),
\end{equation}
where $\Psi_{\text{gp},i}$ is a GP model calibrated from the training set $\left\{\mathbf{x}^{(k)},\Psi_i(\mathbf{x}^{(k)})\right\}_{1\leq k \leq N}$ as detailed in the following. 

As stated by~\citet{rasmussen2006}, a GP is a random process (here the mode $\Psi_i$) indexed over a domain (here $\mathbb{R}^d$), for which any finite collection of process values (here $\left\{\Psi_i(\mathbf{x}^{(k)})\right\}_{1\leq k \leq N}$) has a joint Gaussian distribution. Concretely, let $\widetilde{\Psi}_i$ be a Gaussian random process fully described by its zero mean and its correlation $\pi_i$:
\begin{equation}
\widetilde{\Psi}_i(\mathbf{x})\sim \text{GP}\left(0, \sigma_i^2\,\pi_i(\mathbf{x},\mathbf{x}')\right),
\end{equation}
with $\pi_i(\mathbf{x},\mathbf{x}') = \mathbb{E}\left[\widetilde{\Psi}_i(\mathbf{x})\widetilde{\Psi}_i(\mathbf{x}')\right]$. In our case, the correlation function $\pi$ (or kernel) is chosen as a squared exponential:
\begin{equation}
\pi_i(\mathbf{x}, \mathbf{x}') = \exp\left(-\frac{\|\mathbf{x} - \mathbf{x}'\|^2}{2\,\ell_i^2}\right),
\end{equation}
where $\ell_i$ is a length scale describing dependencies of model output between two input vectors $\mathbf{x}$ and $\mathbf{x}'$, and where $\sigma_i^2$ is the variance of the output signal. Then the surrogate model of interest is the mean of the GP resulting of conditioning $\widetilde{\Psi}_i$ by the training set $\left\{\Psi_i\left(\mathbf{x}^{(k)}\right)\right\}_{1\leq k \leq N}$. For any $\mathbf{x}^*\in\mathbb{R}^d$,
\begin{equation}
\Psi_{\text{gp},i}(\mathbf{x}^*)=\sum_{k = 1}^N\,\beta_{k,i}\,\pi_i\left(\mathbf{x}^*,\mathbf{x}^{(k)}\right),
\end{equation}
where $\beta_{k,i} = \left(\mathbf{\Pi}_i +\tau^2\,\mathbf{I}_N\right)^{-1}\left(\Psi_i\left(\mathbf{x}^{(1)}\right)\ldots \Psi_i\left(\mathbf{x}^{(N)}\right)\right)^T$ with $\mathbf{\Pi}_i=\left(\pi_i\left(\mathbf{x}^{(j)},\mathbf{x}^{(k)}\right)\right)_{1\leq j,k \leq N}$, and where $\tau$ (referred to as the nugget effect) avoids ill-conditioning issues for the matrix $\mathbf{\Pi}$. The hyperparameters $\left\{\ell_i, \sigma_i, \tau\right\}$ are optimized by maximum likelihood applied to the data set $\mathcal{D}_N$ using a basin hopping technique~\citep{wales1997}.

\subsection{Statistical Analysis}\label{sec:SA}

In the following, the water level at curvilinear abscissa $a$ estimated with either the PC surrogate or the pGP surrogate is denoted for simplicity purpose by $\widehat{h}_a$, while $h_a$ denotes the water level obtained using the MASCARET forward model.

\subsubsection{Statistical moments and PDF}

Using a standard MC approach on the reference sampling $\mathcal{D}_{N_{\text{ref}}}$, the water level mean value and standard deviation (STD) at a given curvilinear abscissa $a$ noted $(\mu_{h_a}, \sigma_{h_a})$ are formulated as:
\begin{eqnarray}
\mu_{h_a} &=& \frac{1}{N_{\text{ref}}}\,\displaystyle\sum_{k = 1}^{N_{\text{ref}}}\,h^{(k)}_a, \label{ref:mean_mc} \\
\sigma_{h_a} &=& \sqrt{\frac{1}{N_{\text{ref}}-1}\,\displaystyle\sum_{k = 1}^{N_{\text{ref}}}\,\left(h^{(k)}_a - \mu_{h_a}\right)^2}. \label{ref:std_mc}
\end{eqnarray}
The covariance matrix of the simulated water level denoted by $\mathbf{C} \in \mathcal{M}_M(\mathbb{R})$ is stochastically formulated as:
\begin{equation}
\mathbf{C} = 
\frac{1}{N_{\text{ref}}-1}\,\sum_{k = 1}^{N_{\text{ref}}}\,
\left(\mathbf{h}^{(k)} - \overline{\mathbf{h}}\right)\left(\mathbf{h}^{(k)} - \overline{\mathbf{h}}\right)^T,
\label{eq:cov_mc}
\end{equation}
with $\mathbf{h}^{(k)} = \mathcal{M}(\mathbf{x}^{(k)})$ the $k$th sample in $\mathcal{D}_{N_{\text{ref}}}$ containing the water level at the $M$ observed curvilinear abscissas and with $\overline{\mathbf{h}} = N_{\text{ref}}^{-1}\sum_{k=1}^{N_{\text{ref}}}\mathbf{h}^{(k)}$ the ensemble mean.
 
Using the PC surrogate method, the statistical moments can be derived analytically from the coefficients $\lbrace \gamma_{a,i} \rbrace_{i\in\mathbb{N}^d\atop |i|<P}$ such that at a given curvilinear abscissa $a$, the water level mean and STD ($\mu_{h_{pc,a}}$, $\sigma_{h_{\text{pc},a}}$) read:
\begin{eqnarray}
\mu_{h_{\text{pc},a}} &=& \gamma_{a,0}, \\
\sigma_{h_{\text{pc},a}} &=& \sqrt{\displaystyle\sum_{i\in\mathbb{N}^d\atop |i|<P, i \neq 0}\,\gamma_{a,i}^2}. \label{eq:pcstd}
\end{eqnarray}
The covariance matrix $\mathbf{C}_{\text{pc}}$ can also be directly computed from the coefficients: 
\begin{equation}
\mathbf{C}_{\text{pc}} = \left(\text{cov}\left(h_{pc,a_m},h_{pc,a_n}\right)\right)_{1\leq m,n \leq M }, \quad \text{cov}\left(h_{pc,a_m},h_{pc,a_n}\right)  = \displaystyle\sum_{i\in\mathbb{N}^d\atop |i|<P, i \neq 0}\,\gamma_{a_m,i}\,\gamma_{a_n,i},
\label{eq:covPC}
\end{equation}
with $\text{cov}\left(h_{pc,a_m},h_{pc,a_n}\right)$ the matrix component corresponding to the water level covariance at grid points $a_m$ and $a_n$.

Using the pGP surrogate method, both statistical moments and covariances are stochastically estimated as in the MC approach, meaning that $h$ is replaced by the surrogate $h_{\text{pgp}}$ in Eqs.~(\ref{ref:mean_mc})--(\ref{eq:cov_mc}).

For the MC random sampling method as well as for the PC and pGP surrogate methods, the PDF of the water level at each of the $M = 14$ stations is reconstructed using a kernel smoothing procedure on a large enough stochastic sampling of $\mathcal{M}$, $\mathcal{M}_{\text{pc}}$ or $\mathcal{M}_{\text{gp}}$, respectively~\citep{wand1995,hastie2009}.

\subsubsection{Sensitivity Analysis}
\label{sec:methodo_sa}

Sobol' indices~\citep{sobol1993,saltelli2007} are commonly used for SA based on variance analysis. They provide the quantification of how much of the variance of the quantity of interest is due to the uncertainty in the input parameters assuming these random variables are independent and the random output is squared integrable. The water level (simulated either by MASCARET $h_a$ or by the PC or pGP surrogate $\widehat{h}_a$) variance at curvilinear abscissa $a$ decomposes as: 
\begin{equation}
\mathbb{V}[\widehat{h}_a] = \sum_{i = 1}^{d}\,\mathbb{V}_i(\widehat{h}_a) + \sum_{j=i+1}^{d}\mathbb{V}_{ij}(\widehat{h}_a) + \cdots + \mathbb{V}_{1,2,...,d}(\widehat{h}_a),
\end{equation}
where $\mathbb{V}_i(\widehat{h}_a)= \mathbb{V}\left[\mathbb{E}(\widehat{h}_a|x_i)\right]$, $\mathbb{V}_{ij}(\widehat{h}_a)= \mathbb{V}\left[\mathbb{E}(\widehat{h}_a|x_i x_j)\right] - \mathbb{V}_i(\widehat{h}_a) - \mathbb{V}_j(\widehat{h}_a)$ and more generally, for any $I\subset\{1,\ldots,d\}$, $\mathbb{V}_{I}(\widehat{h}_a)=\mathbb{V}\left[\mathbb{E}(\widehat{h}_a|x_I)\right]-\sum_{J\subset I\,\text{s.t.}\,J\neq I} \mathbb{V}_{J}(\widehat{h}_a)$.
The Sobol' indices for water level at curvilinear abscissa $a$ with respect to the $i$th and $j$th components of the random input vector read: 
\begin{align}
S_i^a = \frac{\mathbb{V}_i(\widehat{h}_a)}{\mathbb{V}(\widehat{h}_a)},\qquad S_{ij}^a = \frac{\mathbb{V}_{ij}(\widehat{h}_a) - \mathbb{V}_i(\widehat{h}_a) - \mathbb{V}_j(\widehat{h}_a)}{\mathbb{V}(\widehat{h}_a)}.
\end{align}
$S_{i}^a$ is the first order Sobol' index corresponding to the ratio of output variance due to the $i$th input parameter uniquely, and $S_{ij}$ is the second-order Sobol' index describing the ratio of output variance due to the $i$th parameter in interaction with the $j$th parameter. Also the total Sobol' index that corresponds to the whole contribution of the $i$th input parameter reads:
\begin{align}
S_{T_i}^a = \sum_{I\subset\{1,\ldots,d\}\atop I \ni i}\,S_I^a.
\end{align}

For both MC and pGP approaches, the Sobol' indices are stochastically estimated. The conditional variances are estimated using Martinez' formulation. This estimator is stable and it provides asymptotic confidence intervals for first order and total order indices~\citep{baudin2016}.

For the PC approach, Sobol' indices can be directly derived from the PC coefficients. For the $i$th component of the input random variable $\mathbf{x}$, the Sobol' index reads:
\begin{equation}
S_{\text{pc},i}^a = \frac{1}{{\sigma_{h_{\text{pc},a}}^2}}\,\displaystyle\sum_{j \in \mathbb{N}^d~\text{s.t.}~|j|<P,\atop j_i > 0~\text{and}~j_{k\neq i}=0}\gamma_{a,j}^2,
\end{equation} 
with $\sigma_{h_{\text{pc},a}}$ the STD computed in Eq.~\eqref{eq:pcstd}. 

\subsection{Error Metrics}\label{sec:error}

In the present study, three error metrics are used to assess the quality of the surrogate water level at a given curvilinear abscissa $\widehat{h}_a$: the $Q_2$ predictive coefficient~\citep{marrel2009}, the Kolmogorov-Smirnov test to evaluate the similarity between PDF~\citep{clarke1992} and the Root Mean Square Error (RMSE). The validation is carried out over a full database $D_{N_{\text{ref}}}$ of size $N_{\text{ref}}$. 

\subsubsection*{Predictive coefficient $Q_2$}
At a given curvilinear abscissa $a$, the $Q_2$ predictive coefficient reads:
\begin{align} 
Q_{2,a} = 1 - \frac{\displaystyle\sum_{k = 1}^{N_{\text{ref}}}\,\left(h^{(k)}_a - \widehat{h}^{(k)}_a\right)^2}{\displaystyle\sum_{k = 1}^{N_{\text{ref}}}\,\left(h^{(k)}_a - \overline{h}_a\right)^2}.
\end{align}

\subsubsection*{Kolmogorov-Smirnov statistical test}

Let $T_F$ (resp. $T_G$) be a random variable with cumulative distribution function (CDF) $F$ (resp. $G$). Let $F_n$ (resp. $G_m$) be its empirical CDF built from $n$ (resp. $m$) independent realizations of $T_F$ (resp. $T_G$). Then, let us define the test statistics:
\begin{equation}
D = \sup_{x} \lvert F_n(x) - G_m(x)\rvert.
\end{equation}
The null hypothesis for the Kolmogorov-Smirnov statistical test supposes that $T_F$ and $T_G$ are identically distributed, i.e.~$F=G$. The Kolmogorov-Smirnov test leads us to reject this hypothesis with a type I error $\alpha\in]0,1[$ when:
\begin{equation}
D > c(\alpha)\,\sqrt{\frac{n+m}{nm}}, \label{eq:nullhypothesis}
\end{equation}
with $c(\alpha)$ a tabulated value found in the literature~\citep{smirnov1939}. In the present study, using $\alpha = 0.05$ and $n = m = N_{\text{ref}}$, the null hypothesis is rejected if $D > \numprint{6.082e-3}$.

\subsubsection*{Root mean square error (RMSE)}

The RMSE is used to evaluate the accuracy in the estimation of the correlation matrix and of the Sobol' indices with respect to the MC stochastic estimation over the data set $D_{N_{\text{ref}}}$, i.e.
\begin{equation}\label{eq:rmse}
\text{RMSE} = \sqrt{\frac{1}{M}\,
\displaystyle\sum_{a = 1}^{M}\,(\Box_{a} - \Box_{\text{mc},a})^2.}
\end{equation}
where $\Box$ designates one component of the correlation matrix or one Sobol' index at the given curvilinear abscissa $a$.

\subsection{Numerical implementation and validation}
\label{sec:OT_JPOD}

Due to the increasing interest in UQ over the last decade, a significant number of UQ-dedicated tools/libraries are now available for the scientific community. OpenTURNS (see www.openturns.org) is an open-source (GNU LGPL) scientific library developed since 2005 by EDF, Airbus, Phimeca and IMACS and usable as a Python module dedicated to uncertainty treatment and reliability analysis in a structured industrial approach~\citep{baudin2015}. OpenTURNS offers a wide catalogue of features for UQ (e.g.~PC expansion, \citeauthor{dutka2009}~\citeyear{dutka2009}) and benefits from a well-organized developers' and users' community (forum, training, user guides). It can either be used as a Python module or as a component within a coupling platform. For further information on the OpenTURNS library, the reader is referred to the online reference and use case guides.

In the present study, MASCARET and OpenTURNS are both integrated components of the SALOME platform developed at EDF (see www.salome-platform.org). This integrated framework allows for an efficient use of MASCARET as a Python function for simulating an ensemble of water levels and for building the PC surrogate models using the methods implemented in OpenTURNS. The pGP surrogate models rely on the use of the JPOD Python tool developed at CERFACS~\citep{braconnier2011} since 2007 and recently augmented with OpenTURNS' UQ capabilities~\citep{roy2016}. It acts as a Python platform that is able to interact with an external simulation code, i.e.~generate a design of experiment, run the corresponding simulations and perform statistical analysis of the outputs. The GP implementation relies on the package \emph{Scikit-learn}~\citep{pedregosa2011}. 

Both PC and pGP strategies were validated on: (i)~classical optimization functions with scalar output such as Ishigami~\citep{ishigami1990} and Michalewicz~\citep{molga2005}; and (ii)~for a functional output case using simplified 1-D open-channel flow equations (backwater curves)~\citep{elmocaydEMA,roy2016}. JPOD has been successfully tested and validated on industrial configurations during the European SimSAC and ALEF (FP7) projects.

\section{Comparison of Polynomial Chaos (PC) and Gaussian Process (pGP) surrogates}\label{sec:UQresults}

\subsection{Reference Monte Carlo (MC) results}
\label{sec:UQresults_ref}

We first present the results obtained using the MC reference sampling ($N_{\text{ref}} = 100,000$ in $\mathcal{D}_{N_{\text{ref}}}$) in terms of water level mean, STD, PDF, correlation matrix as well as Sobol' indices associated with $Q$ and $K_{s_3}$. These results are used as reference to evaluate the accuracy of the PC and pGP surrogate models.

\begin{figure*}[!h]               
\centering
\subfigure[Along the 50 km reach.]{
\includegraphics[width=0.47\linewidth,height=\textheight,keepaspectratio]{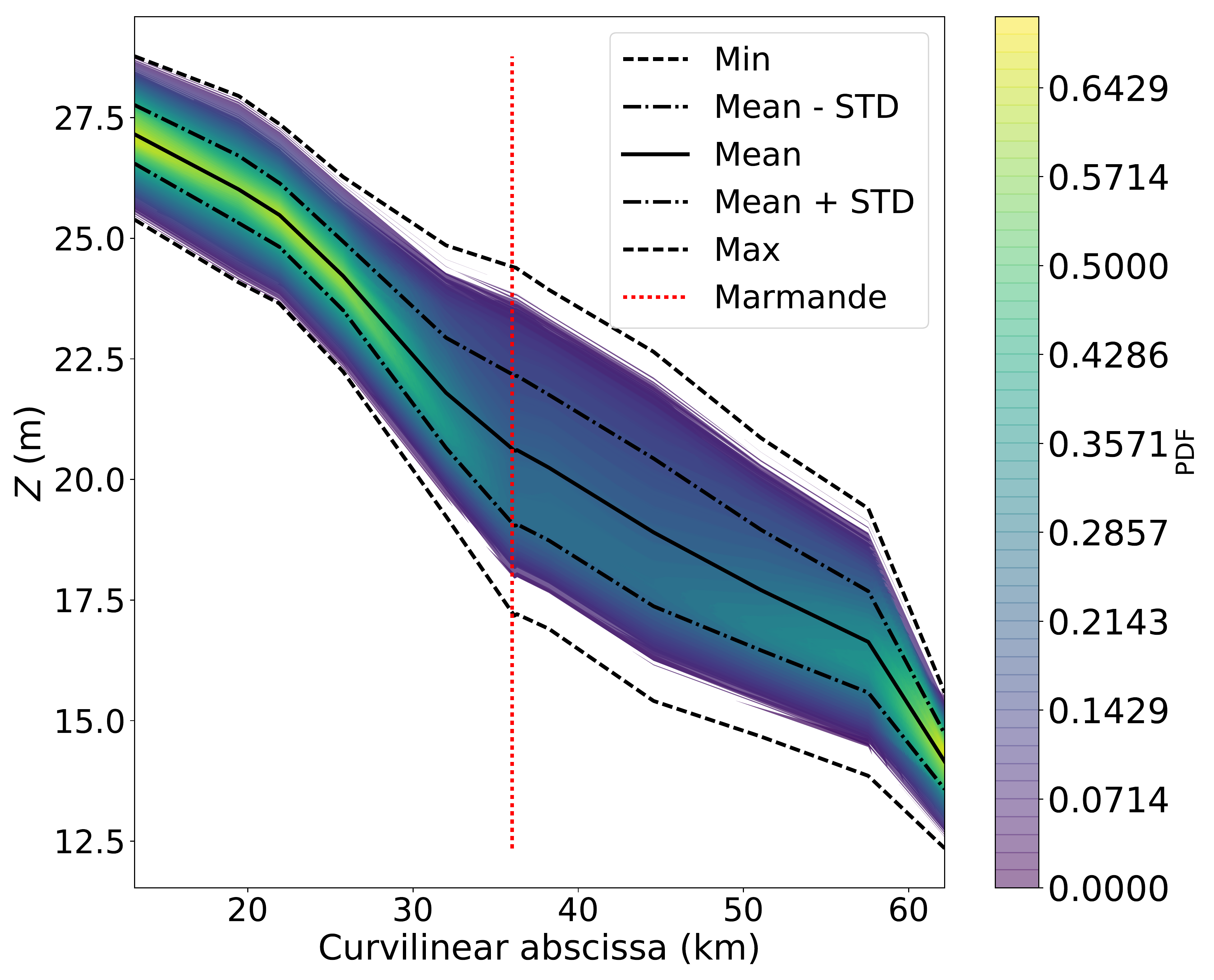}} 
\subfigure[At Marmande ($a = 36$~km).]{
\includegraphics[width=0.47\linewidth,height=\textheight,keepaspectratio]{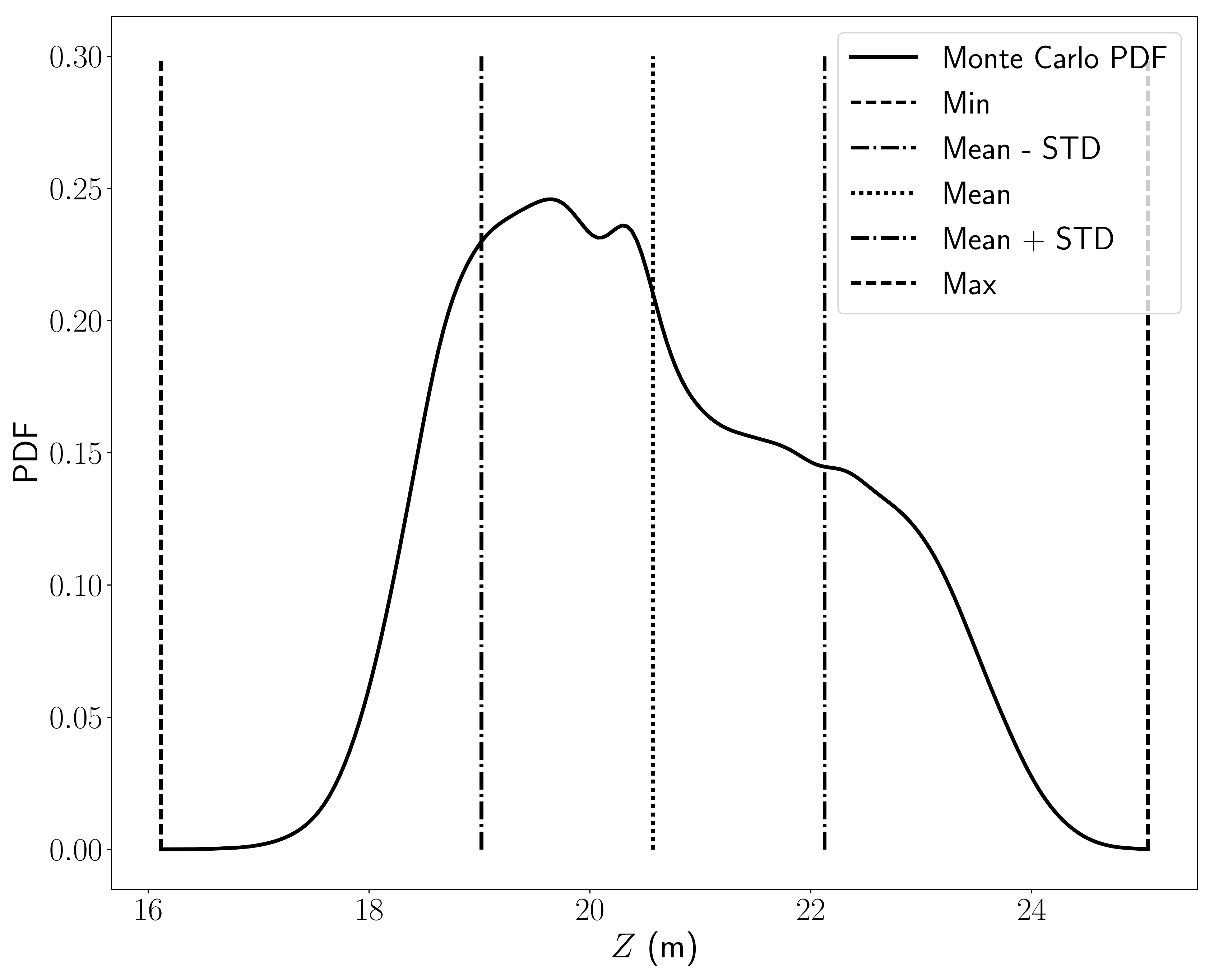}}
\caption{Reference PDF of the water elevation obtained with the $N_{\text{ref}} = 100,000$ snapshots in $\mathcal{D}_{N_{\text{ref}}}$ derived from MC random sampling: (a)~at the $M = 14$ stations along the 50 km river reach; (b)~at Marmande. The solid line indicates the mean water level with respect to the curvilinear abscissa; the dotted lines indicate the spread corresponding to 2~STD; the dashed lines indicate the maximum and minimum water level values; and the vertical dotted line indicates Marmande's location.}
\label{fig:monte-carlo_pdf}
\end{figure*}

Figure~\ref{fig:monte-carlo_pdf}a displays the water level PDF computed from $\mathcal{D}_{N_{\text{ref}}}$ data set integrated with a MC approach for the $M = 14$ stations along the curvilinear abscissa $a \in [a_{\text{in}}, a_{\text{out}}]$. The mean water level (Eq.~\ref{ref:mean_mc}) is represented with a thick black line; the interval between two STD (Eq.~\ref{ref:std_mc}) is represented with dotted lines; and the minimum and maximum water level values are represented with dashed lines. The upstream part of the river is under the influence of the upstream forcing, the spread of the ensemble tends to increase. The downstream part of the river is under the influence of the downstream boundary condition where the water level and discharge are related by the local rating curve $RC$, the spread of the ensemble tends to decrease near the downstream boundary condition. At Marmande ($a = 36$~km), the flow is complex due to strong variation of the local bathymetry (Fig.~\ref{fig:garonne}b), the ensemble spread is larger and the PDF plotted in Fig.~\ref{fig:monte-carlo_pdf}b features two main modes due to the change in backwater curves solutions for subcritical flow. 
 
\begin{figure*}[!h]               
\centering
\subfigure[]{
\includegraphics[width=0.45\linewidth,height=\textheight,keepaspectratio]{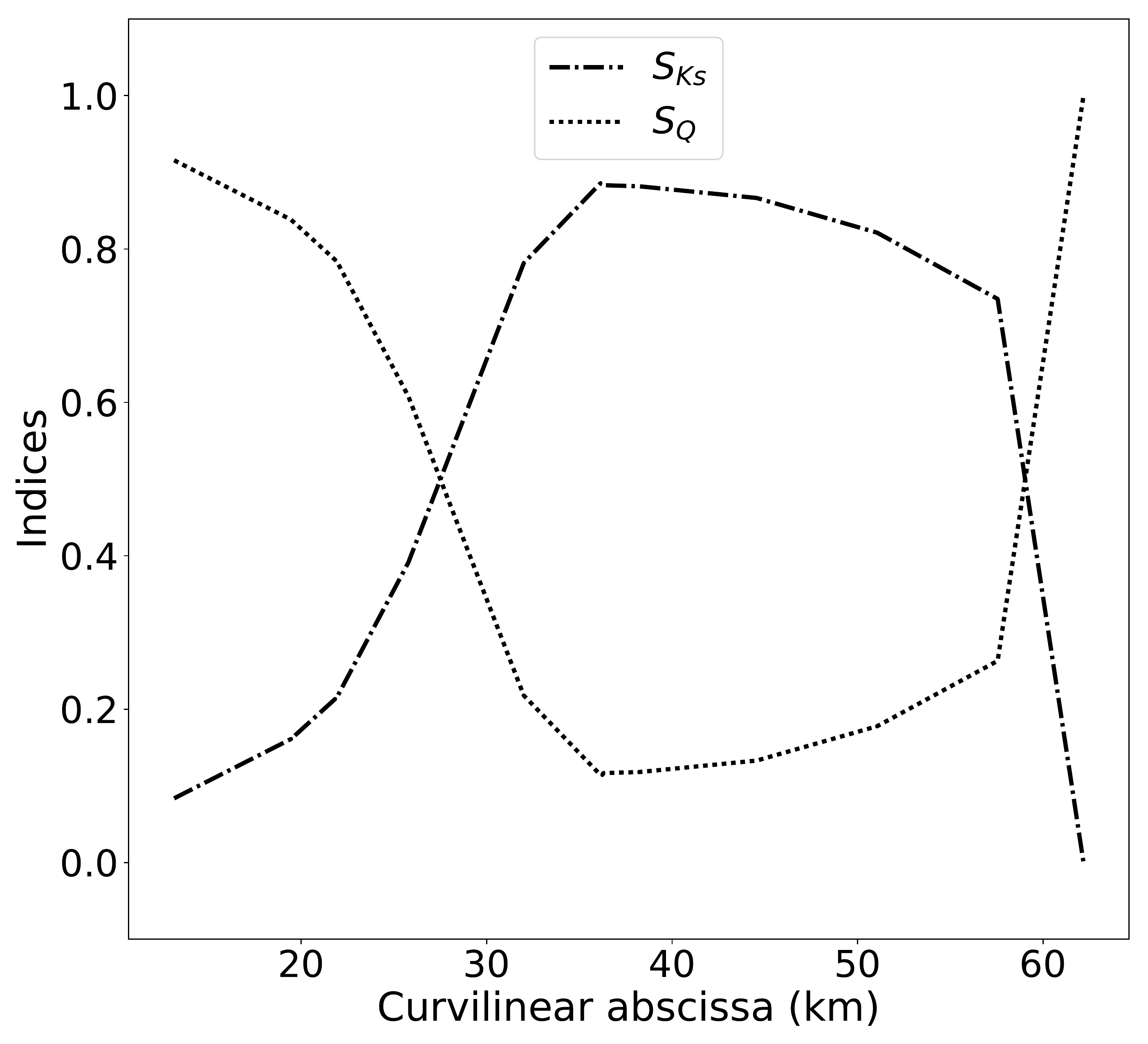}} 
\subfigure[]{
\includegraphics[width=0.45\linewidth,height=\textheight,keepaspectratio]{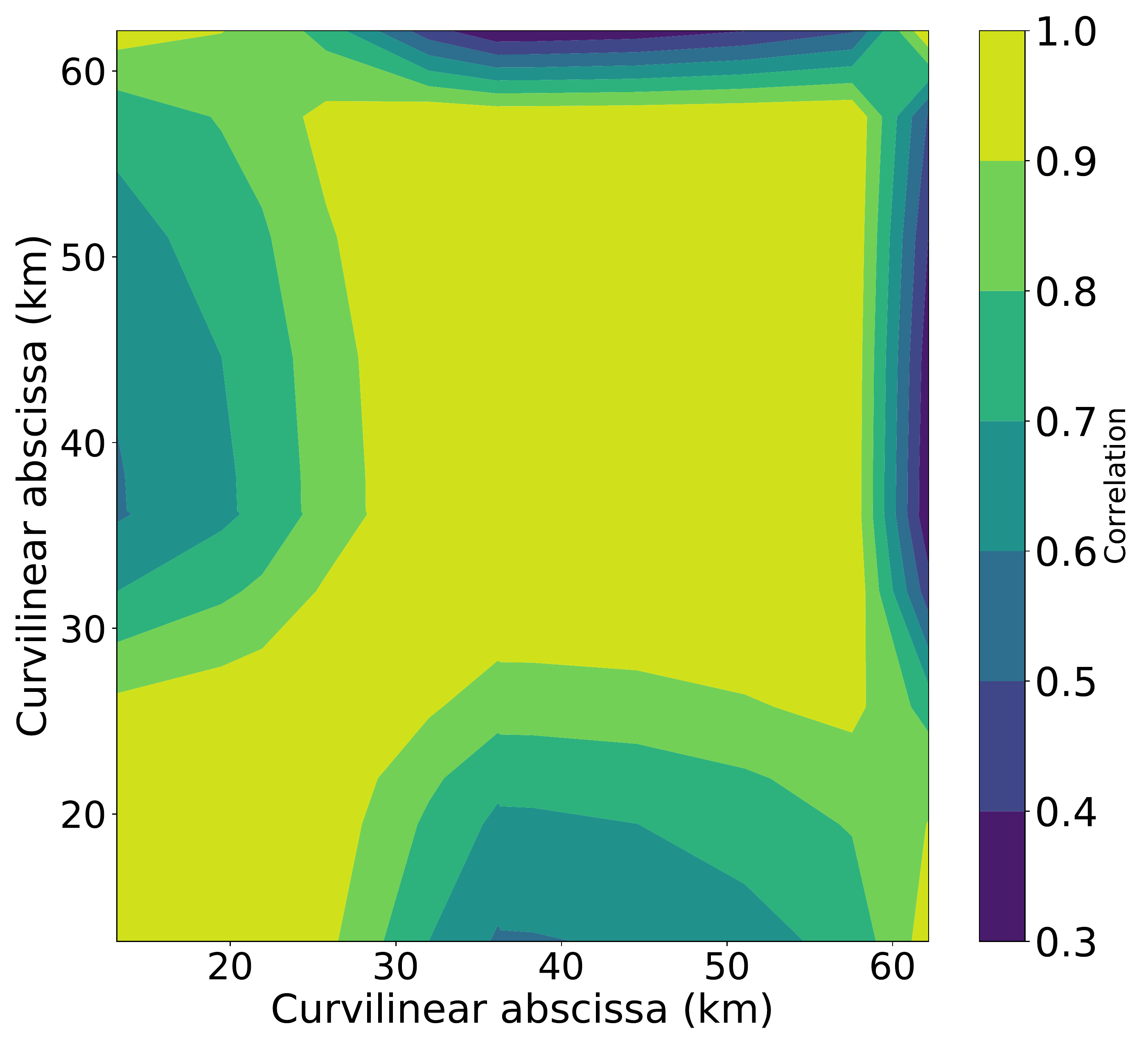}}
\caption{Measures of importance using MC random sampling. (a)~Reference Sobol' first order indices along the 50 km reach. Dashed-dotted line corresponds to the Sobol' index associated with the upstream discharge $Q$. Dotted line corresponds to that associated with the Strickler coefficient $K_{s_3}$. (b)~Reference spatial correlation matrix $\mathbf{C}_{\text{mc}}$ associated with the spatially distributed water level $\mathbf{h}$.}
\label{fig:monte-carlo_sobol_corr}
\end{figure*}

The Sobol' indices for the water level $S_Q$ and $S_{K_{s_3}}$ computed from $\mathcal{D}_{N_{\text{ref}}}$ with respect to $Q$ and $K_{s_3}$ are presented in Fig.~\ref{fig:monte-carlo_sobol_corr}a along the curvilinear abscissa $a$. These indices confirm the previously mentioned spatial sensitivity. The water level variance is mostly explained by the upstream discharge variability for $a = [0; 30~\text{km}]$. It is then mostly explained by the Strickler coefficient variance for $a = [30; 60~\text{km}]$. Near the downstream boundary condition, the water level is related to the discharge by the local rating curve $RC$, the very last part of the network is thus under the influence of the discharge. First and total order indices are equal, meaning that there is no correlation between the errors in the input parameters. 

Figure~\ref{fig:monte-carlo_sobol_corr}b displays the water level correlation matrix along the 50 km reach (derived using Eq.~\ref{eq:cov_mc}) that is estimated from $\mathcal{D}_{N_{\text{ref}}}$. The $n$th column of the matrix describes the water level error correlations between one given location on the channel $a_n$ and the rest of the channel $a_m$ with $a_m \in [a_{\text{in}}, a_{\text{out}}]$. By definition, the correlation is equal to 1 on the diagonal, it decreases when the distance between $a_n$ and $a_m$ increases. We first analyse the correlation function for a point located upstream of the river ($a_i = 15~\text{km}$) where $S_Q = 0.9$ and $S_{K_{s_3}}= 0.1$. Water level errors are strongly correlated in the upstream part of the river, which is under the influence of the upstream discharge boundary condition, where $S_Q$ is large. Errors between $a = 15$~km and the rest of the river tends to decorrelate when the influence of $K_{s_3}$ increases (i.e.~where $S_{K_{s_3}}$ is larger). We then analyse the correlation function for Marmande ($a = 36~\text{km}$) where $S_Q= 0.15$ and $S_{K_{s_3}}= 0.85$. The correlation between water level errors at Marmande and the rest of the river is large in the vicinity of Marmande, where the influence of $K_{s_3}$ prevails. It then decreases for upstream and downstream locations that are under the influence of the upstream discharge and the downstream rating curve $RC$ (where $S_Q$ is large), respectively.

\subsection{Convergence Analysis for Surrogates}

We now use the same metrics as for the MC random sampling approach in Sec.~\ref{sec:UQresults_ref} to evaluate the accuracy of both PC and pGP surrogates. The surrogate models are built using a training set $(\mathcal{X}, \mathcal{Y})$ of size $N$ that is much smaller than that of the reference sample $N_{\text{ref}}$; they are then validated with respect to the reference MC results. The PC surrogate is built using a Gaussian quadrature rule with $N = (P+1)^2$ particles in the training set ($P$ is the total polynomial degree to be determined). The pGP approach is built using an approximate low-discrepancy Halton's sequence of the same budget as for the PC approach. This is approximate in the sense that we consider the closest values to the standard Halton's sequence that are part of the data set $\mathcal{D}_{N_{\text{ref}}}$. The sensitivity to the value of $P$ and thus to the size of the training set $N$ is investigated.

\begin{figure}[H]               
\centering
\subfigure[pGP - 49 snapshots.]{
\includegraphics[width=0.47\linewidth,height=\textheight,keepaspectratio]{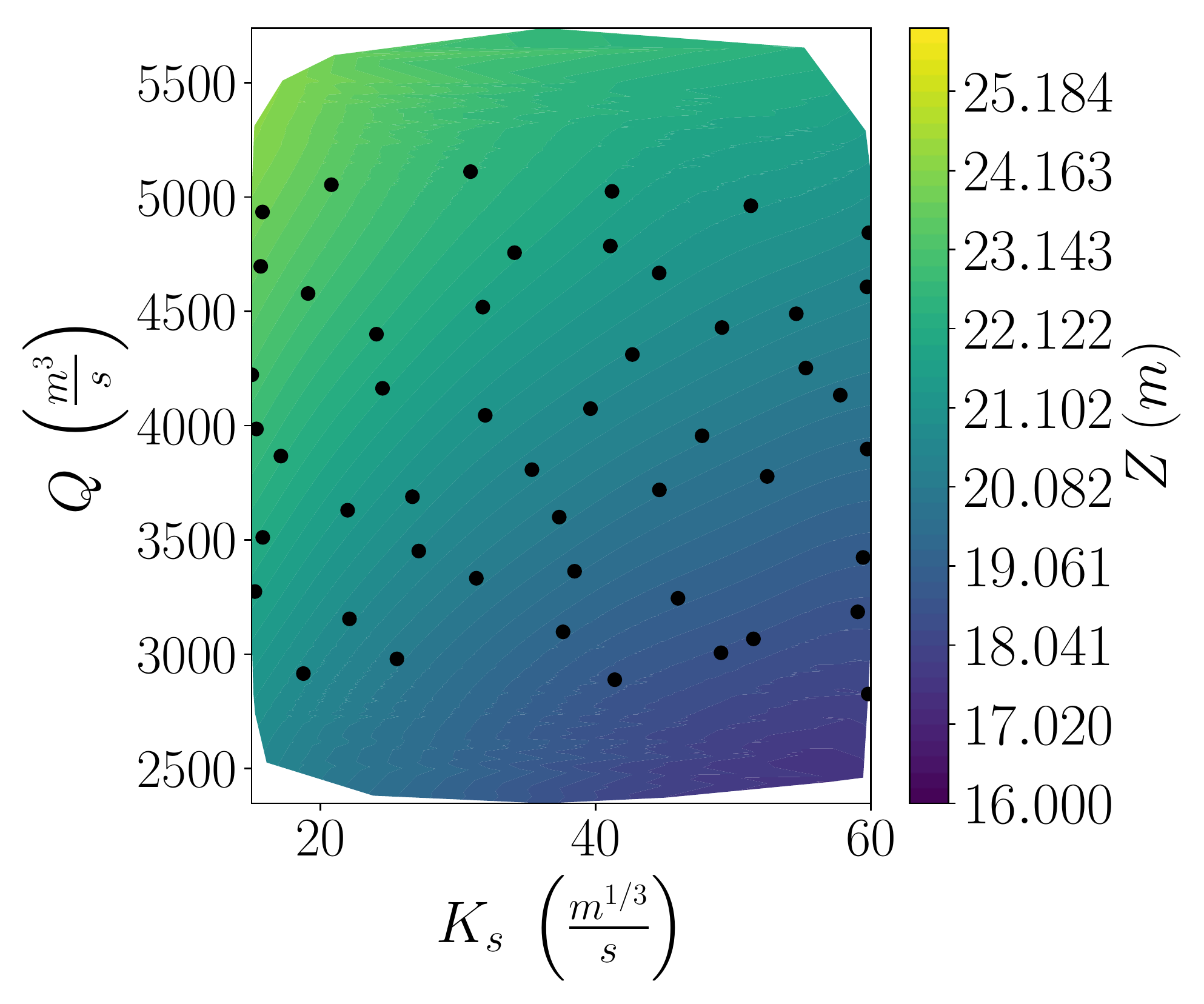}}          
\subfigure[pGP - 121 snapshots.]{
\includegraphics[width=0.47\linewidth,height=\textheight,keepaspectratio]{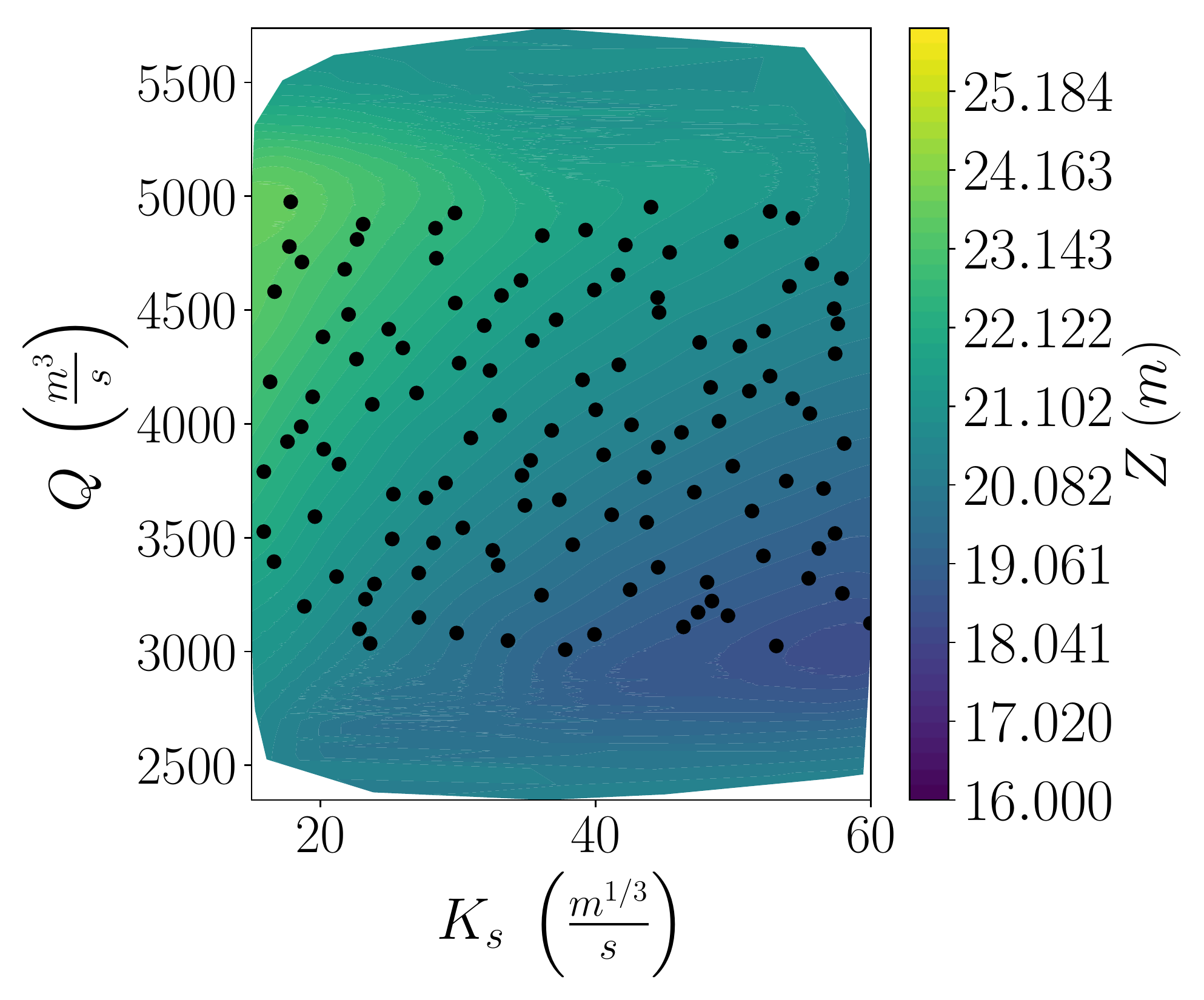}}
\subfigure[PC - 49 snapshots.]{
\includegraphics[width=0.47\linewidth,height=\textheight,keepaspectratio]{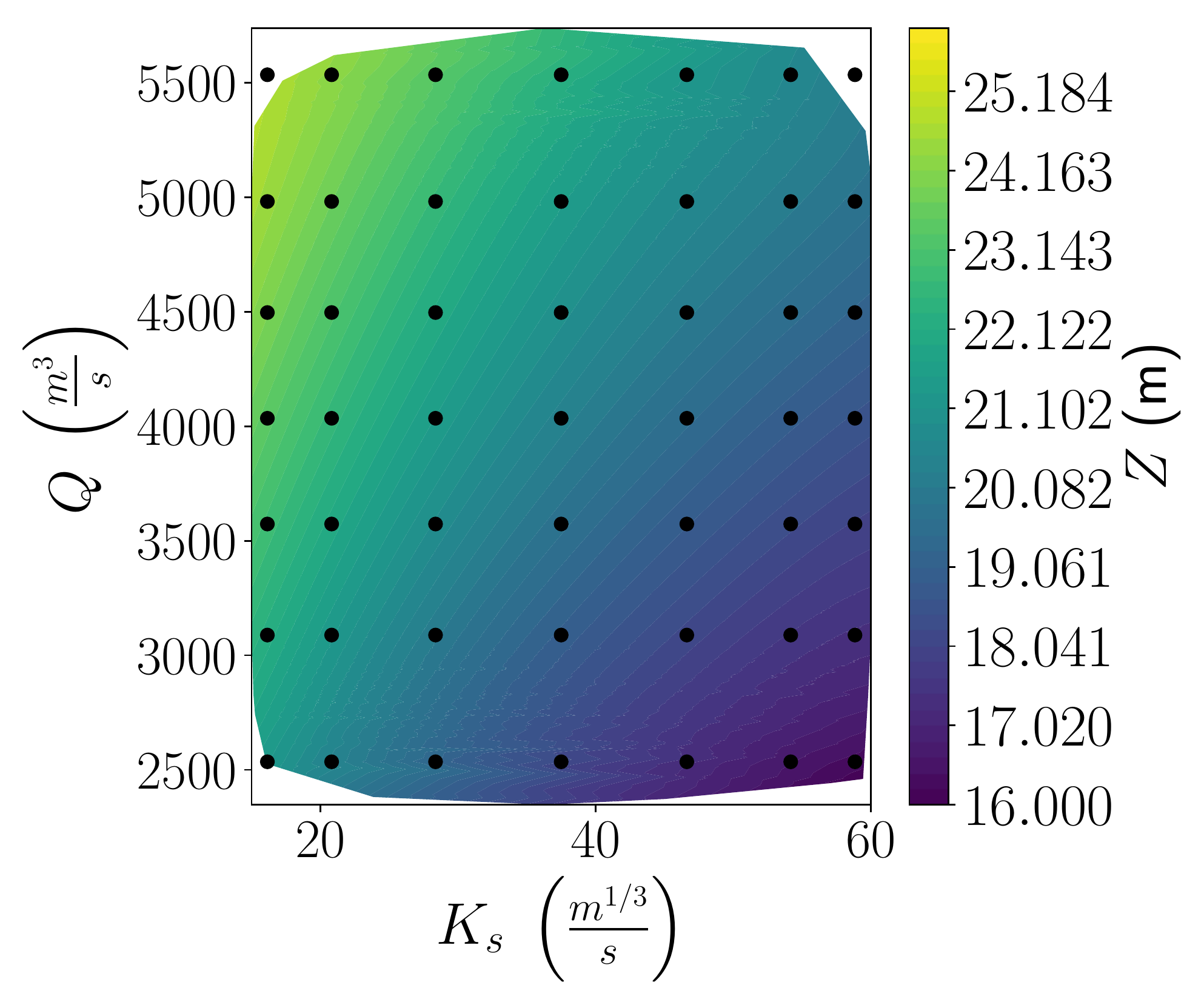}}           
\subfigure[PC - 121 snapshots.]{
\includegraphics[width=0.47\linewidth,height=\textheight,keepaspectratio]{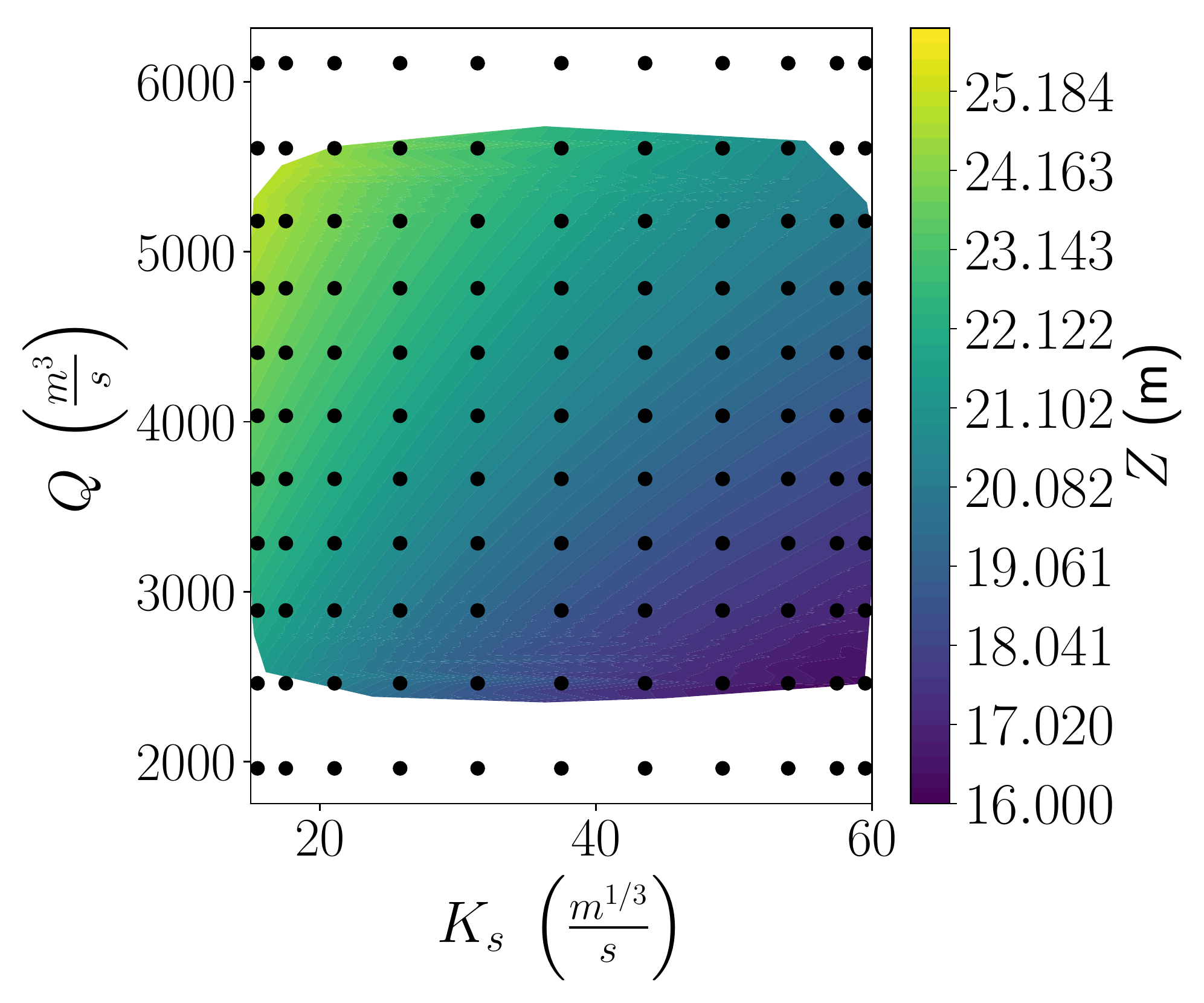}}
\caption{Water level response surface at Marmande computed at $\mathcal{D}_{N_{\text{ref}}}$. Top: pGP using (a)~$N = 49$ snapshots and (b)~$N = 121$ snapshots. Bottom~: PC using (c)~$N=49$ snapshots and (d)~$N=121$ snapshots. Black dots represent the design of experiments used to construct the surrogate models. The colour map corresponds to the evaluation of the surrogate over the full data set $\mathcal{D}_{N_{\text{ref}}}$.{\color[rgb]{0.9,0.2,0.16}\protect\footnotemark}}
\label{fig:response-surface-pgp-pc}
\end{figure}

Both surrogates are computed with a fixed budget $N$ of 49 and 121 MASCARET evaluations. For PC, this value of $N$ corresponds respectively to $P = 6$ and $P = 10$. The pGP and PC response surfaces at Marmande are presented in Fig.~\ref{fig:response-surface-pgp-pc}. The design of experiment is represented by black dots. The colour map is evaluated by sampling each surrogate over the full data set $\mathcal{D}_{N_{\text{ref}}}$ made of $N_{\text{ref}} = 100,000$ particles. It is found that the water level increases with increasing discharge $Q$ and decreasing Strickler coefficient $K_{s_3}$, consistently with MASCARET behaviour. Due to the quadrature rule, increasing the number of snapshots (from $P = 6$ in Fig.~\ref{fig:response-surface-pgp-pc}c to $P = 10$ in Fig.~\ref{fig:response-surface-pgp-pc}d) allows to build a higher order PC surrogate valid on a wider input range for $Q$ that is described by a Gaussian PDF. For $P = 10$, some of the quadrature roots are outside of the MC sample and require additional MASCARET evaluations to build the PC surrogate $\mathcal{M}_pc$. Looking at the pGP design of experiments, the input space interval for $Q$ has been arbitrarily fixed to optimally represent the PDF. Since we consider a Gaussian distribution, its range has been bounded to $[3000;5000~\unit{m^3\,s^{-1}}]$. Following Chebyshev's theorem, this leads to a~90~\% confidence interval.

The distribution of the particles in the design of experiments used by PC and pGP differs. This is done on purpose based on previous performance tests carried out with respect to the training set size $N$ and evaluating the surrogate accuracy, see~\citep{elmocaydEMA} and~\citep{roy2016}, respectively. On the one hand, the design of experiments for the PC surrogate is constrained by the PDF of the uncertain inputs. We use here a quadrature-based PC since it was found to be more cost-effective than regression-based PC for a small dimensional problem ($d = 2$) on the Garonne case~\citep{elmocaydEMA}. On the other hand, using the approximate Halton's sequence is known to be accurate for pGP surrogate~\citep{damblin2013}. This is indeed useful to cover the uncertain space without any bias and to have low discrepancy, meaning that most of the quantity of interest variance is captured and that a good $Q_2$ criterion is achieved. The choice of the design of experiment will be driven in future work according to the study objectives. For instance, DA usually relies on the assumption of Gaussian PDF for the input parameters, so the design of experiments can use this prior information. For risk analysis, threshold values are paramount and a design of experiment accounting for parameter space extrema is required.

\footnotetext{Scaling of (d) was set in order to show the entire design of experiments. This allows to grasp the fact that the parameter space is evaluated in regions that are not evaluated by the Monte-Carlo sampling.}

The $Q_2$ error between the surrogate water level and the forward model water level for $D_{N_{\text{ref}}}$ averaged over the river is given in Table~\ref{tab:valid_model} for different  sizes of training set $N$ varying between 49 and 256. The error remains below $\numprint{e-3}$, even for $N = 49$ snapshots; it is only slightly improved when the number of snapshots $N$ increases to 256 and beyond. 
\begin{table}[H]
\centering
\caption{$Q_2$ error for pGP and PC surrogates computed with respect to the MC reference. The error corresponds to the average over the $M$ stations with increasing number of snapshots $N$ from 49 to 256.}
\begin{tabular}{lcc}
\toprule
$N$ & pGP & PC \\
\cmidrule{2-3}
49  & 0.99965 & 0.99983\\
121 & 0.99514 & 0.99993\\
256 & 0.99143 & 0.99962\\
\bottomrule
\end{tabular}
\label{tab:valid_model}
\end{table}

\begin{figure*}[!h]               
\centering
\subfigure[pGP -- $a = 15~\text{km}$]{
\includegraphics[width=0.47\linewidth,height=\textheight,keepaspectratio]{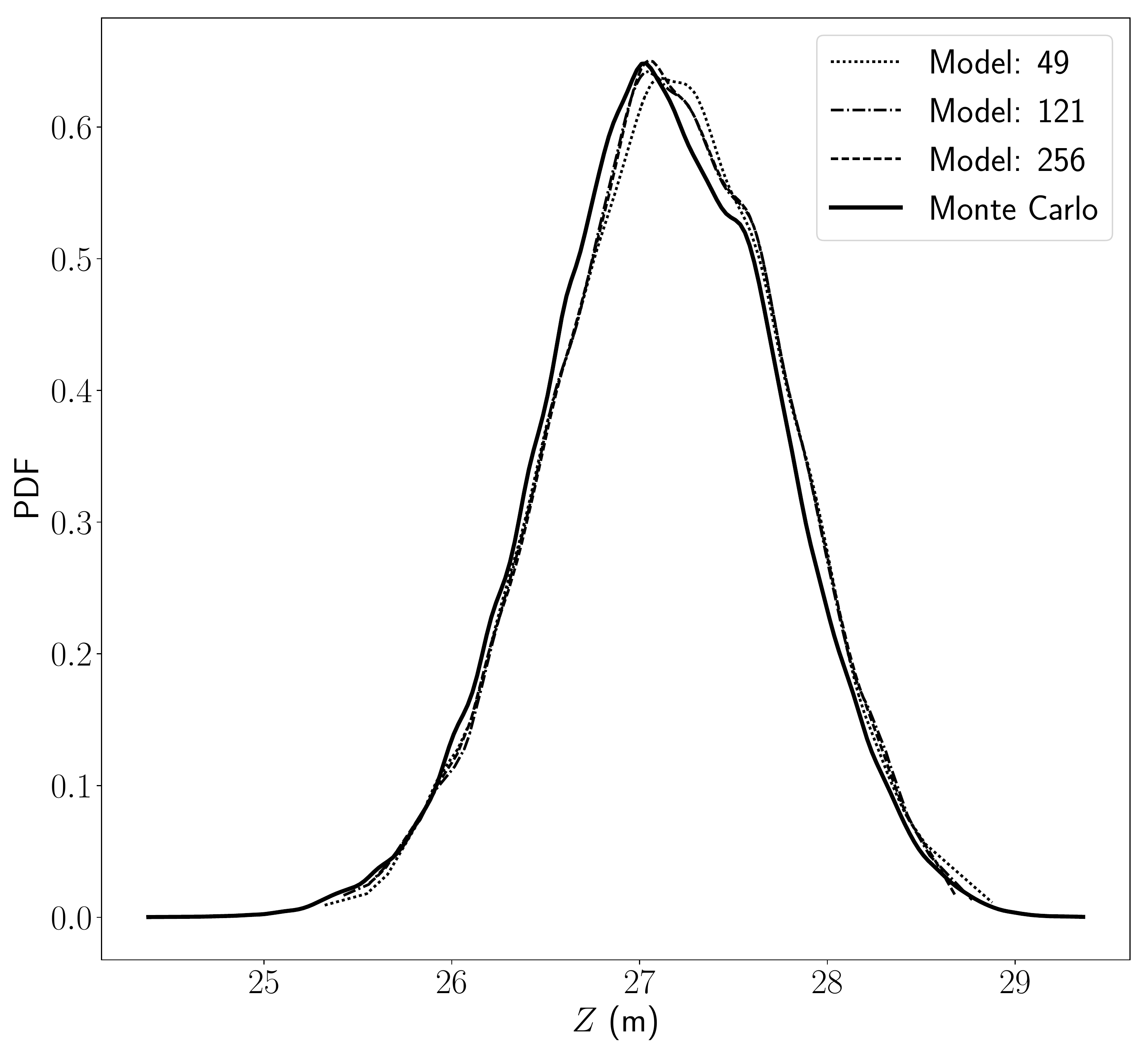}}
\subfigure[pGP -- $a = 36~\text{km}$]{
\includegraphics[width=0.47\linewidth,height=\textheight,keepaspectratio]{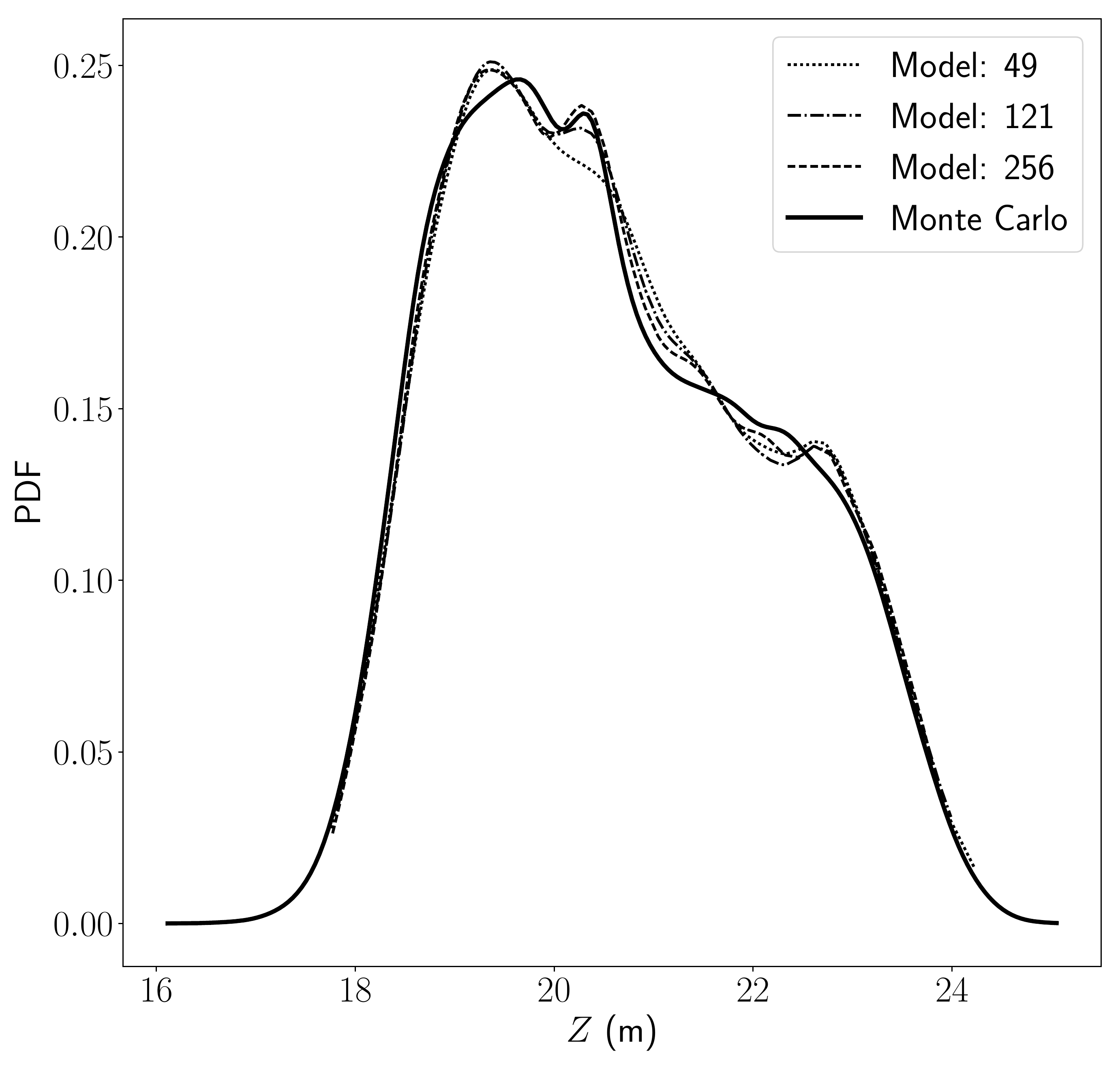}}
\subfigure[PC -- $a = 15~\text{km}$]{
\includegraphics[width=0.47\linewidth,height=\textheight,keepaspectratio]{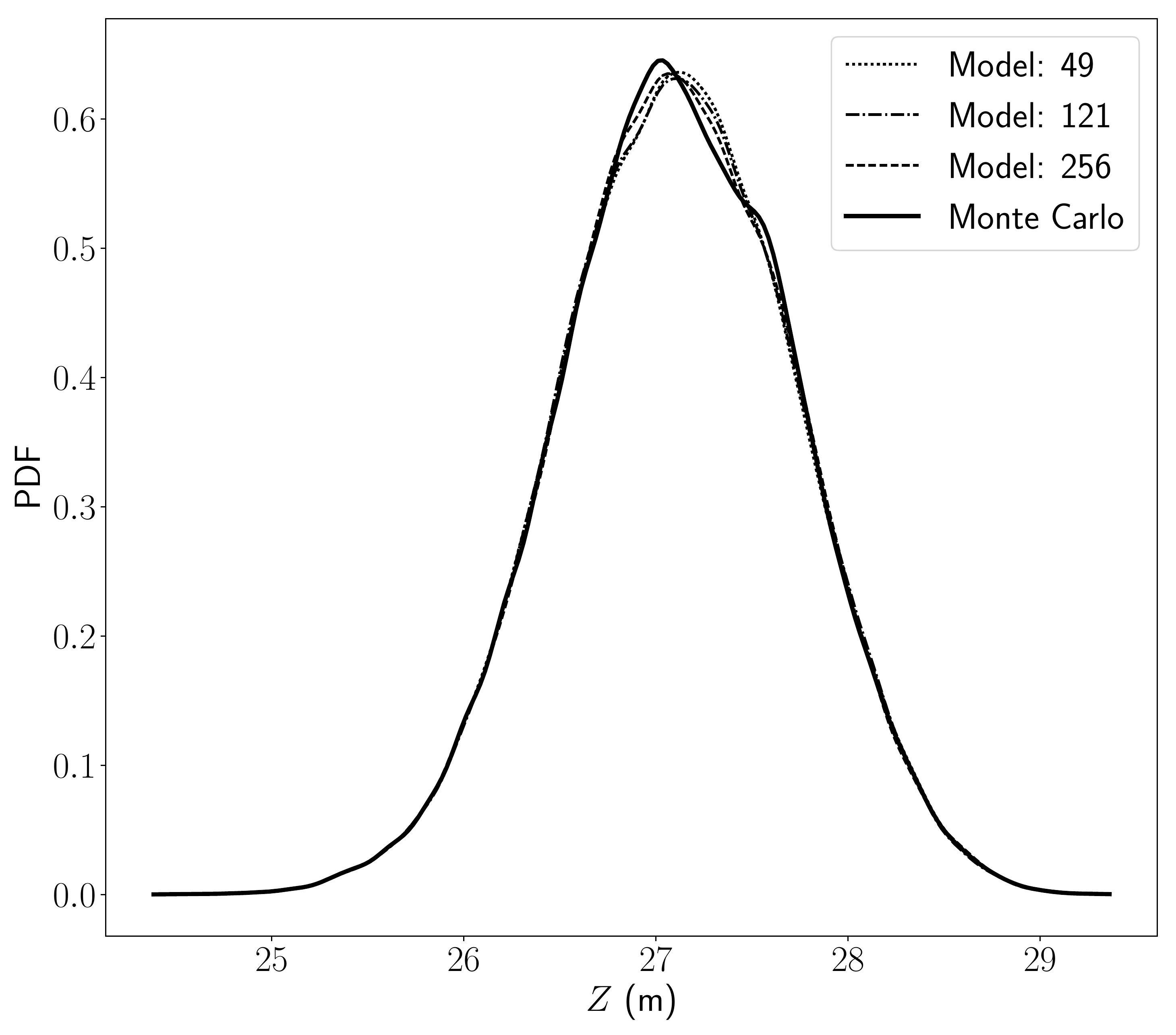}}
\subfigure[PC -- $a = 36~\text{km}$]{
\includegraphics[width=0.47\linewidth,height=\textheight,keepaspectratio]{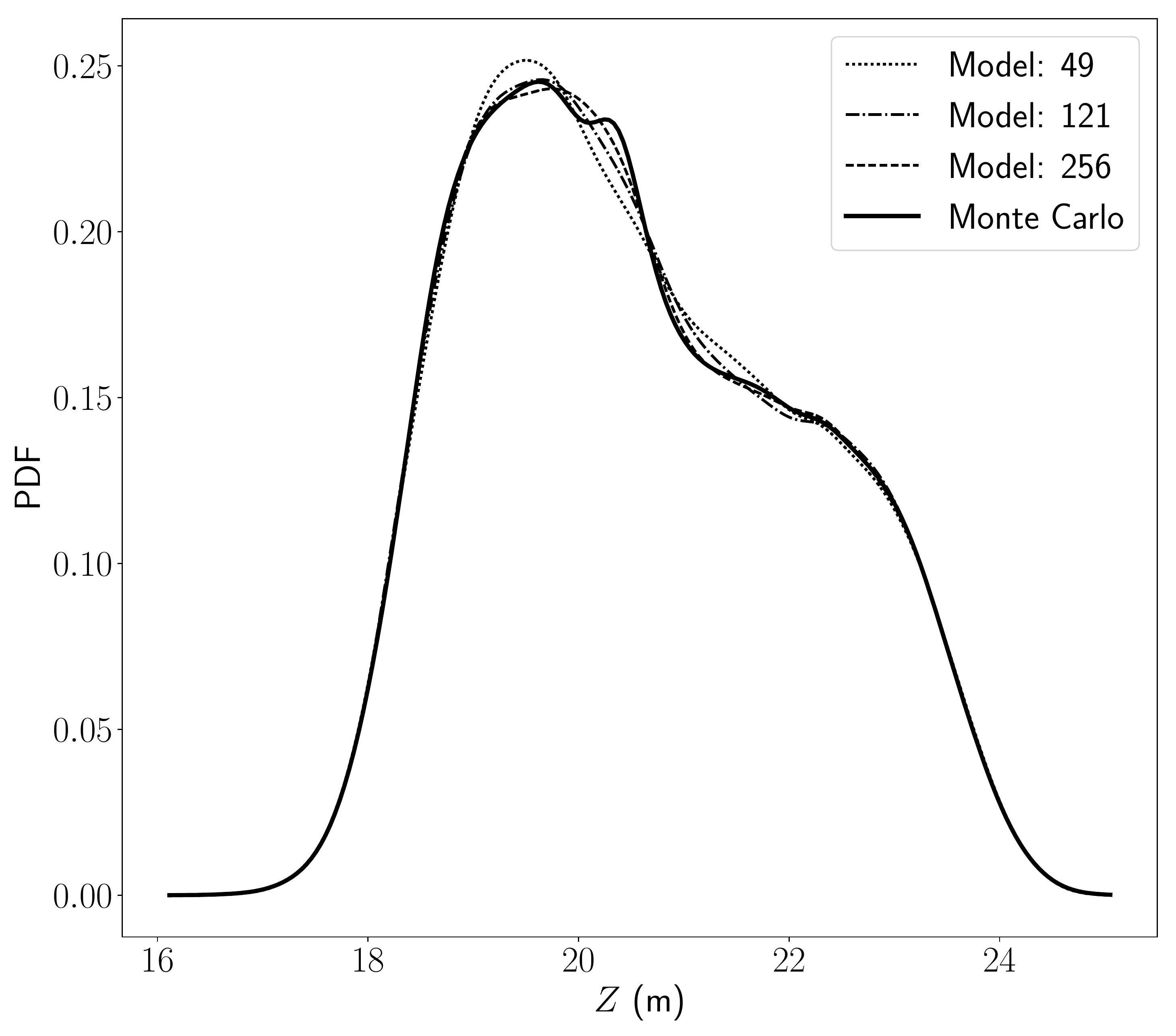}}
\caption{Comparison of water level PDF obtained with pGP (top panel) and PC (bottom panel) (a)--(c)~At $a = 15~\text{km}$ (near upstream boundary condition). (b)--(d)~At $a=36~\text{km}$ (Marmande). The comparison is carried out for different sizes of training set $N$ (49, 121, 256); the MC result is provided as reference in solid black line.}
\label{fig:pdf-station-0_9}
\end{figure*}

The water level PDF estimated with the PC and the pGP surrogate models based on $49, 121, 256$ snapshots are compared in Fig.~\ref{fig:pdf-station-0_9} at the curvilinear abscissa $a = 15~\text{km}$ (near upstream boundary condition) and $a = 36~\text{km}$ (Marmande). In the upstream part of the river, the PDF is uni-modal and is well represented with a small number of snapshots for both surrogates. On the contrary, at Marmande, the dynamics of the flow is more complex and the PDF is bimodal. Both PC and pGP surrogates are able to retrieve the overall shape of the PDF at $a = 15~\text{km}$ and $a = 36~\text{km}$. The shape is more accurate when the number of snapshots $N$ increases. This is quantified using a Kolmogorov-Smirnov statistical test, which measures the fit between the water level CDF computed from each surrogate model and that computed from the reference MC. Table~\ref{tab:ks} indicates that the null hypothesis (Eq.~\ref{eq:nullhypothesis}) is rejected for both surrogates computed with 49 snapshots (when $D > \numprint{6.082e-3}$) and accepted when at least 121 snapshots are used. For $N = 49$, Fig.~\ref{fig:pdf-station-0_9}b--d show that the location of the first mode is shifted compared to MC reference. Increasing $N$ to 121 enables the PC approach to correctly locate this mode, while it enables the pGP approach to represent the second mode with an accurate amplitude, leading to an accepted null hypothesis. Each approach presents a particular limitation: the first mode is not well positioned by the pGP approach; the second mode is not captured by the PC approach. As for the tail of the PDF, it is correctly represented by the PC surrogate, while the pGP surrogate slightly oscillates around the shape of the reference MC PDF.

\begin{table}[H]
\centering
\caption{Two-sample Kolmogorov-Smirnov statistical test for pGP and PC surrogates with respect to the MC reference at Marmande with increasing number of snapshots $N$ (49, 121, 256). The null hypothesis is rejected if $D > \numprint{6.082e-3}$.}
\begin{tabular}{llcc}
\toprule
Surrogate & Snapshots & Statistics \emph{D} & \emph{p}-value \\
\cmidrule{3-4}
&49  & $\numprint{7.95e-3}$ & 0.004 \\
pGP&121 & $\numprint{3.97e-3}$ & 0.409 \\
&256 & $\numprint{3.02e-3}$& 0.751\\
\cmidrule{3-4}
&49  & $\numprint{7.15e-3}$ & 0.012 \\
PC&121 & $\numprint{4.95e-3}$ & 0.172  \\
&256 & $\numprint{4.93e-3}$ & 0.175\\
\bottomrule
\end{tabular}
\label{tab:ks}
\end{table}

\begin{figure*}[!h]               
\centering
\subfigure[pGP -- 121 snapshots]{
\includegraphics[width=0.47\linewidth,height=\textheight,keepaspectratio]{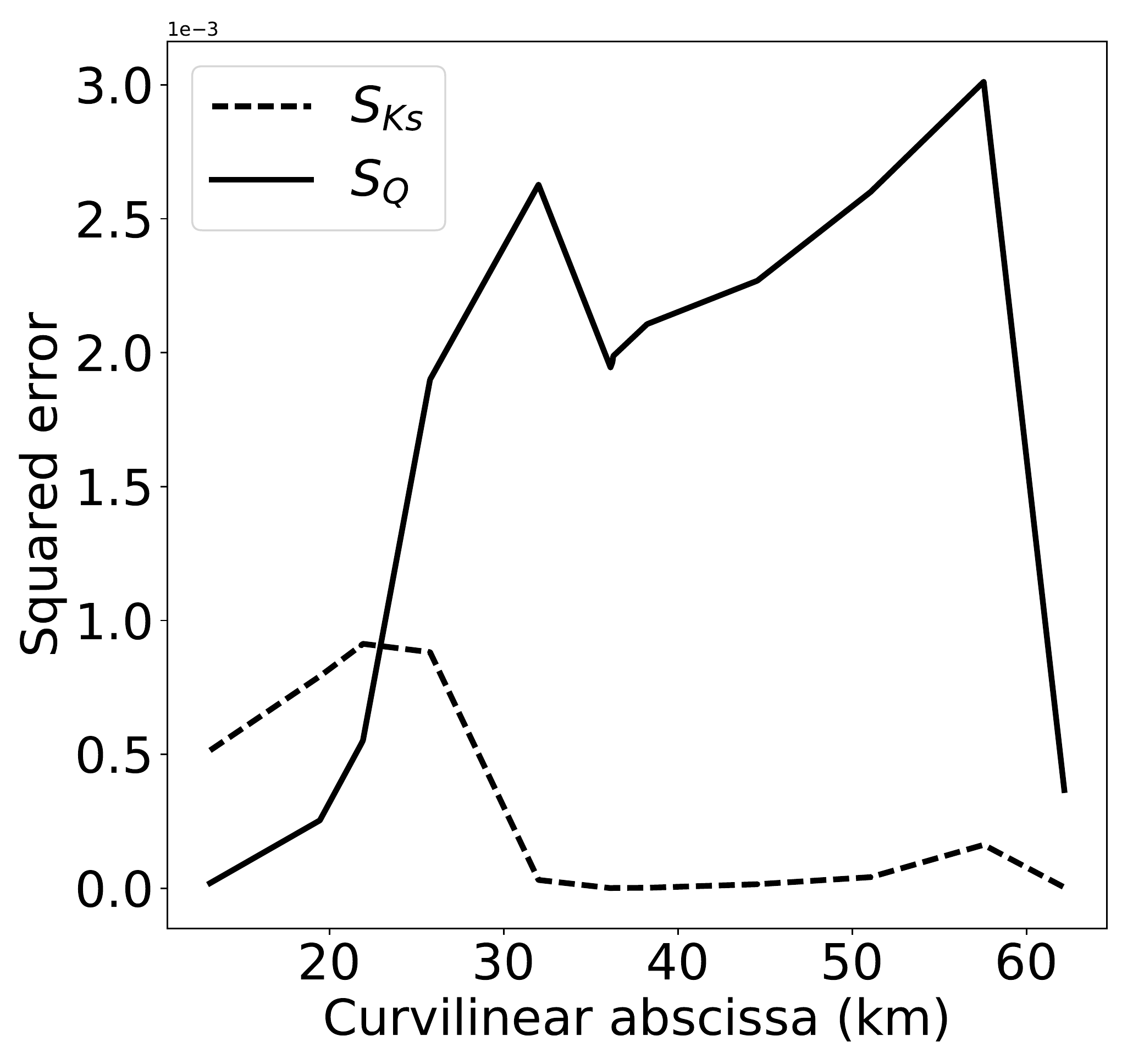}}
\subfigure[PC -- 121 snapshots]{
\includegraphics[width=0.47\linewidth,height=\textheight,keepaspectratio]{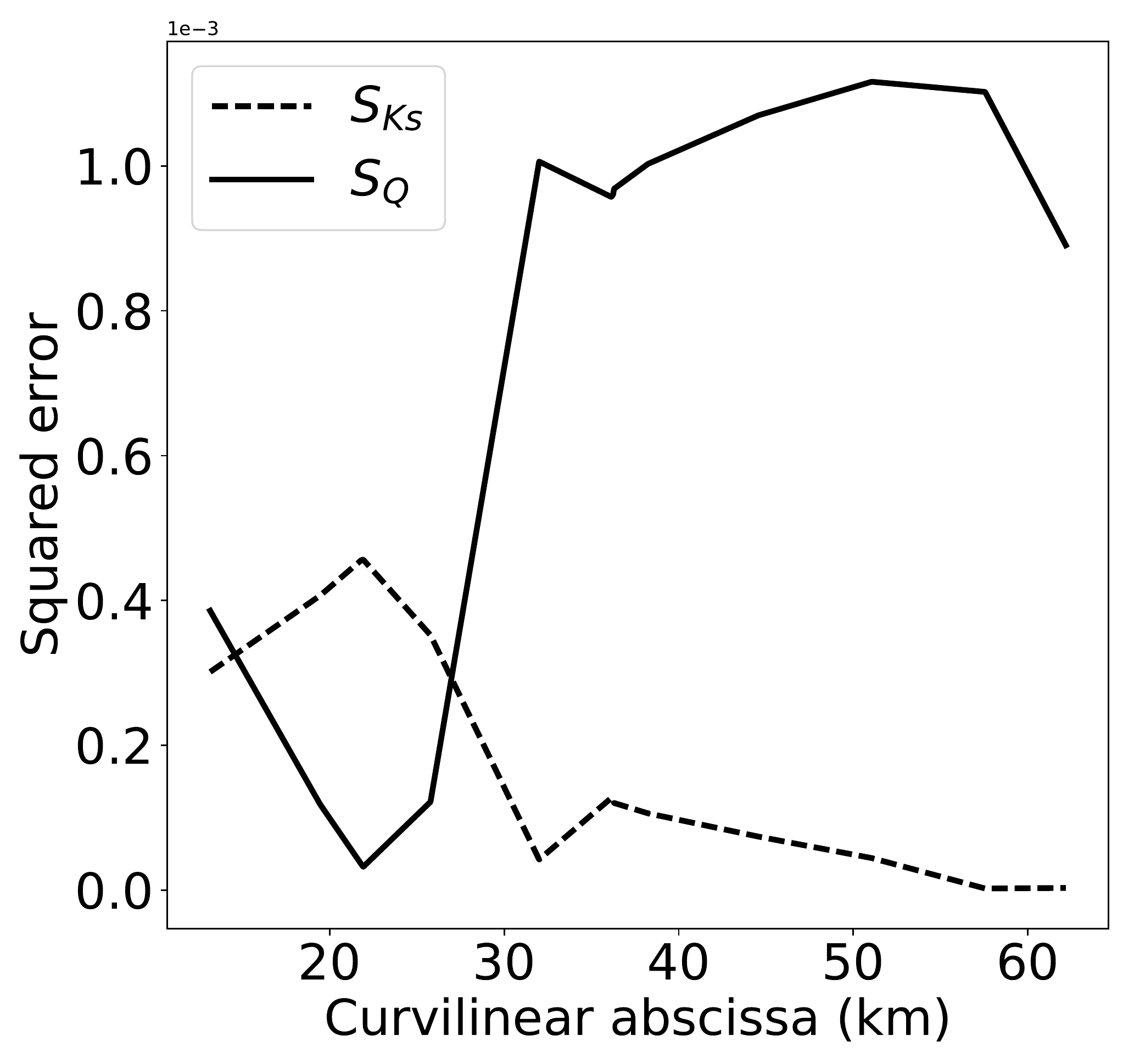}}
\caption{Squared error of the spatial Sobol' indices along the 50 km reach for (a)~pGP and (b)~PC surrogate models built using $N = 121$ snapshots in the training set. Dashed lines correspond to the Sobol' index associated with $K_{s_3}$; solid lines correspond to that associated with the upstream discharge $Q$.}
\label{fig:sobol-map}
\end{figure*}

\begin{figure*}[!h]               
\centering
\subfigure[pGP -- 121 snapshots]{
\includegraphics[width=0.47\linewidth,height=\textheight,keepaspectratio]{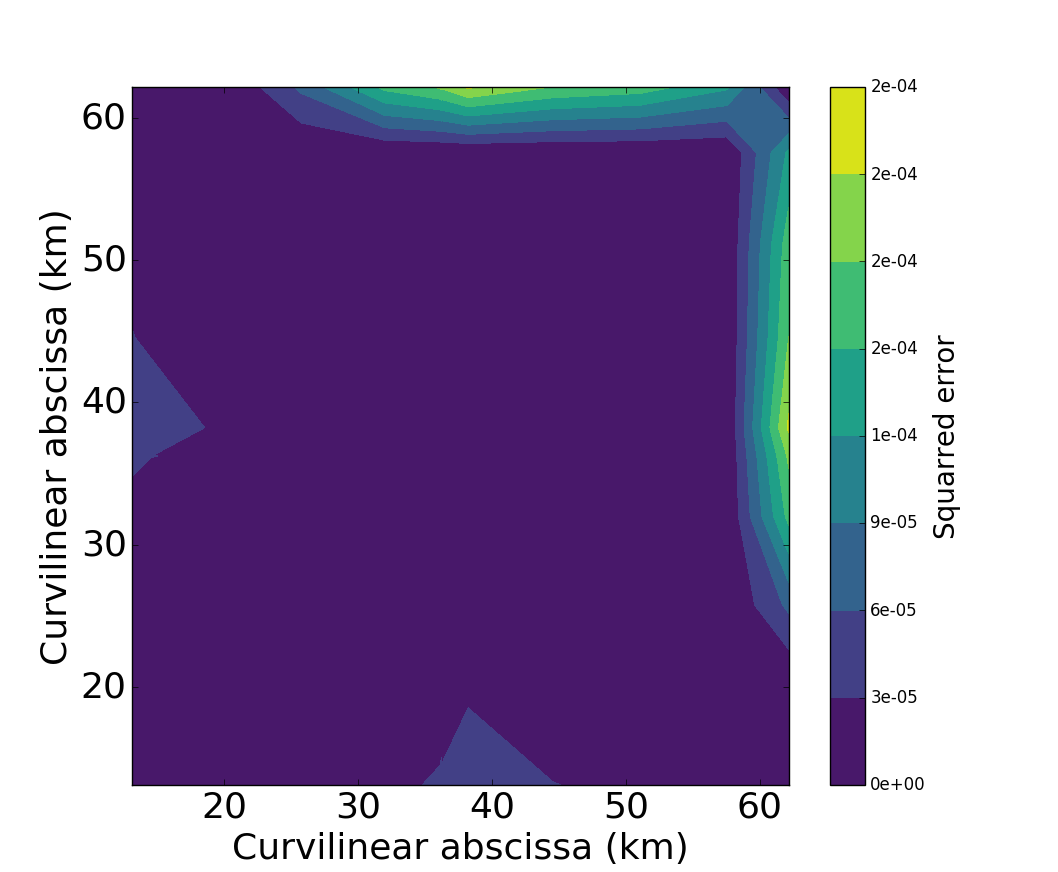}}
 ~ 
\subfigure[PC -- 121 snapshots]{
\includegraphics[width=0.47\linewidth,height=\textheight,keepaspectratio]{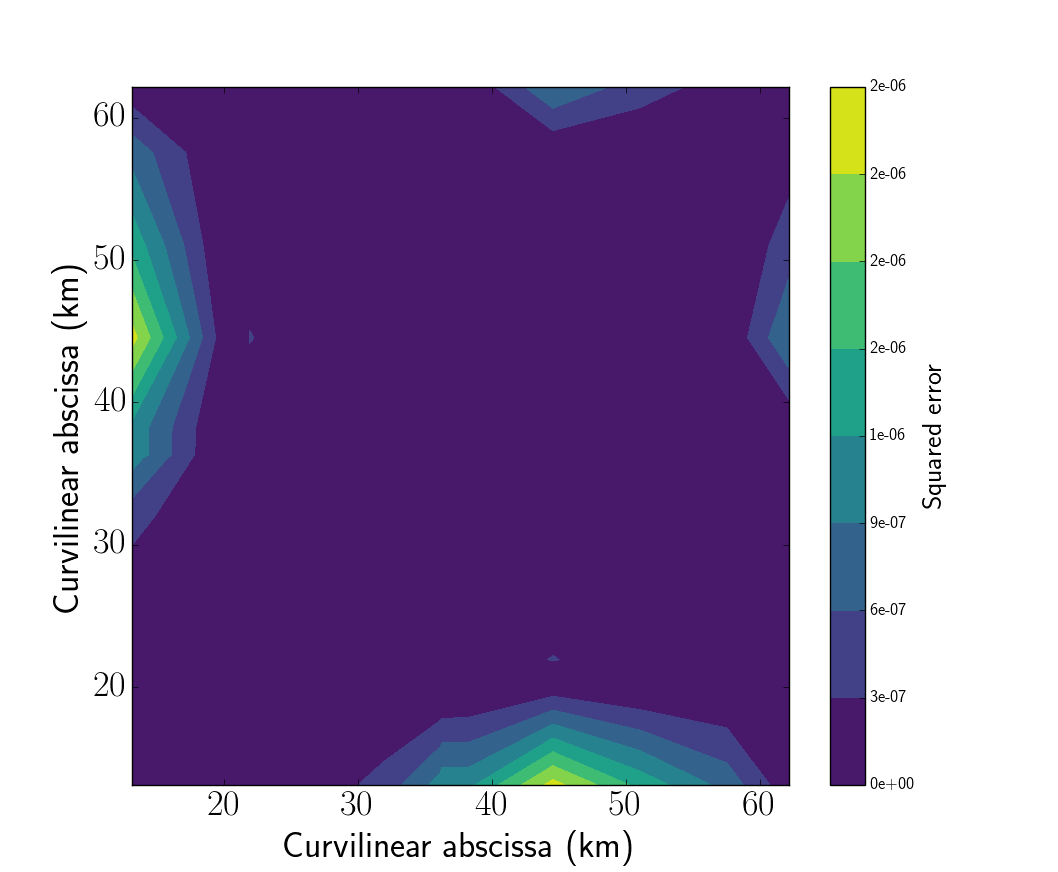}}
\caption{Squared error of the correlation matrix for (a)~pGP and (b)~PC surrogate models built using $N = 121$ snapshots.}
\label{fig:corr-mse}
\end{figure*}

The RMSE for the Sobol' indices for both pGP and PC with a fixed computational budget of $N = 121$ simulations is of the order of $10^{-2}$ when computed over the 50 km reach. Figure~\ref{fig:sobol-map} displays the squared error for the Sobol' indices along the curvilinear abscissa (Eq.~\ref{eq:rmse}); the spatial pattern is similar for both surrogates and the squared error for each index is larger where the Sobol' indices are larger. The RMSE for the correlation matrix for both pGP and PC with $N = 121$ simulations are equal to $\text{RMSE}_{\text{pc}} = \numprint{3.67e-4}$ and $\text{RMSE}_{\text{gp}} = \numprint{4.59e-3}$. The spatial distribution of the squared error is plotted in Fig.~\ref{fig:corr-mse}. These results confirm the good behaviour of both PC and pGP surrogates with respect to MASCARET. For both Sobol' indices and correlation matrices, the PC surrogate slightly outperforms pGP. 


\section{Discussion and Conclusions}\label{sec:ccl}

The purpose of this work was to compare two popular strategies for building surrogate models, Polynomial Chaos (PC) and POD-based Gaussian Process (pGP). Both methods were applied to a hydraulic case corresponding to a spatially distributed open-channel steady flow along the Garonne River depending on the upstream discharge $Q$ and on the Strickler friction coefficient $K_{s_3}$, with the long-term objective to determine which surrogate strategy could be useful in the framework of ensemble-based data assimilation. It is important to show how surrogate models could be used to estimate some statistical quantities in a cost-effective way. This is useful for sensitivity analysis studies to evaluate the impact of physical parameters and external forcing on the river state. This is also useful for data assimilation to estimate correlation matrix and PDF related to the spatially distributed river state for the Ensemble Kalman Filter (EnKF) and the Particle Filter (PF), respectively.

We carried out a convergence study based on the following metrics: water level statistical moments, correlation matrix, PDF as well as Sobol' indices representing the contribution of the upstream discharge and the Strickler friction coefficient on the water level variance. The accuracy of the PC and pGP surrogates were measured by their ability to retrieve the reference metrics obtained with a converged Monte Carlo (MC) random sample (including $\numprint{100,000}$ MASCARET simulations). An in-depth comparison of these metrics was done using the same computational budget: 121 MASCARET snapshots. The sensitivity to the number of snapshots was carried out to ensure that 121 simulations were enough to make this comparison valuable.

Results showed that both surrogate models can be used in place of the MASCARET forward model for uncertainty propagation without loss of accuracy. None of the two surrogate models clearly outperforms the other. In both cases, the PC and pGP surrogate models are able to correctly retrieve all physical information. The PC strategy seems to be more precise to compute the spatially distributed correlations as well as the Sobol' indices, with the advantage that these indices do not need any further MASCARET evaluation as they are analytically computed from the PC expansion. Still, the multimodal water level PDF at Marmande (which is an important observation station along the Garonne River in operational context) was better captured by the pGP strategy that requires an additional sampling of the surrogate. Indeed, even increasing the number of snapshots to 256 and beyond was not enough to retrieve the second mode of the PDF using the PC surrogate model, while this was already captured with 121 snapshots by the pGP surrogate model. Still, it should be mentioned that the PC model better positions the first mode than the pGP model when using the MC approach as reference. Last but not least, the PC strategy requires some insight about the uncertain inputs of the system. We may not have access to this information in practice, leading potentially to a poor robustness of the PC surrogate. The quantity of interest may also feature non-linearity and exhibit extrema, which could be difficult to account for using quadrature points that could miss some physics. Alternative (for instance sparse) projection strategies could be investigated in the future to overcome these limitations. 

Conclusions for the present test case highlight the validity of both quadrature-based PC and POD-based GP surrogate strategies for SWE in permanent flow for a small dimensional problem (the size of the uncertain space is $d = 2$). The ranking between PC and pGP approaches will need to be further investigated when moving to open-channel unsteady flow modelling used for instance in the context of operational flood forecasting. The first challenge lies in the extension of the PC and pGP surrogates to larger uncertain dimension $d$, especially to address parameters that vary in space or over time or both such as the bathymetry spatial field and the time series of the upstream discharge. This may require the evaluation of more advanced strategies to reduce the size of the basis, the uncertain dimension (for instance through the Karhunen-Loève transformation) and the number of snapshots $N$ in the training set. For instance, the quadrature method used to build the PC surrogate is known to suffer from the \emph{curse-of-dimensionality}. This will have to be revisited for a larger size $d$ of the uncertain space. The second challenge lies in the validity of the surrogate model over successive data assimilation cycles. It may be necessary to adjust the coefficients of the surrogates to track the changes in the river state behaviour over time. Those are crucial steps for the complementary use of surrogate models within data assimilation algorithms.

\bibliographystyle{plainnat}
\bibliography{biblio_paper.bib}  

\begin{thebibliography}{64}
\providecommand{\natexlab}[1]{#1}
\providecommand{\url}[1]{\texttt{#1}}
\expandafter\ifx\csname urlstyle\endcsname\relax
  \providecommand{\doi}[1]{doi: #1}\else
  \providecommand{\doi}{doi: \begingroup \urlstyle{rm}\Url}\fi

\bibitem[Barth\'el\'emy(2015)]{barthelemy2015}
S.~Barth\'el\'emy.
\newblock \emph{Assimilation de donn\'ee ensembliste et couplage de mod\`eles
  hydrauliques 1D-2D pour la pr\'evision des crues en temps r\'eel: application
  au r\'eseau hydraulique Adour Maritime.}
\newblock PhD thesis, Institut National Polytechnique de Toulouse, 2015.

\bibitem[Baudin et~al.(2015)Baudin, Dutfoy, Iooss, and Popelin]{baudin2015}
Michael Baudin, Anne Dutfoy, Bertrand Iooss, and Anne-Laure Popelin.
\newblock {OpenTURNS: An industrial software for uncertainty quantification in
  simulation}.
\newblock 2015.

\bibitem[Baudin et~al.(2016)Baudin, Boumhaout, Delage, Iooss, and
  Martinez]{baudin2016}
Michael Baudin, Khalid Boumhaout, Thibault Delage, Bertrand Iooss, and
  Jean-Marc Martinez.
\newblock {Numerical stability of Sobol' indices estimation formula}.
\newblock In \emph{8th International Conference on Sensitivity Analysis of
  Model Output,}, R{\'{e}}union Island, 2016.

\bibitem[Berveiller(2005)]{berveiller2005}
Marc Berveiller.
\newblock \emph{El\'ements finis stochastiques : approches intrusive et non
  intrusive pour des analyses de fiabilit\'e}.
\newblock PhD thesis, Universit\'e Blaise Pascal, Clermont-Ferrand, 2005.

\bibitem[Besnard and Goutal(2011)]{besnard2011}
A.~Besnard and N.~Goutal.
\newblock {Comparaison de mod\`{e}les 1D \`{a} casiers et 2D pour la
  mod\'{e}lisation hydraulique d'une plaine d'inondation--Cas de la Garonne
  entre Tonneins et La R\'{e}ole}.
\newblock \emph{La Houille Blanche}, 3:\penalty0 42--47, 2011.

\bibitem[Birolleau et~al.(2014)Birolleau, Poette, and Lucor]{birolleau2014}
A.~Birolleau, G.~Poette, and D.~Lucor.
\newblock {Adaptive Bayesian inference for discontinuous inverse problems,
  application to hyperbolic conservation laws}.
\newblock \emph{Commun. Comput. Phys.}, 16:\penalty0 1--34, 2014.

\bibitem[Blatman(2009)]{blatman2009phd}
G.~Blatman.
\newblock \emph{{Adaptative sparse Polynomial Chaos expansions for uncertainty
  propagation and sensitivity analysis}}.
\newblock PhD thesis, Universit\'e Blaise Pascal, Clermont-Ferrand, 2009.

\bibitem[Bozzi et~al.(2014)Bozzi, Passoni, Bernardara, Goutal, and
  Arnaud]{bozzi2015}
S.~Bozzi, G.~Passoni, P.~Bernardara, N.~Goutal, and A.~Arnaud.
\newblock {Roughness and Discharge Uncertainty in 1D Water Level Calculations}.
\newblock \emph{Environmental Modeling \& Assessment}, 4, 2014.
\newblock \doi{10.1007/s10666-014-9430-6}.

\bibitem[Braconnier et~al.(2011)Braconnier, Ferrier, Jouhaud, Montagnac, and
  Sagaut]{braconnier2011}
T.~Braconnier, M.~Ferrier, J.-C. Jouhaud, M.~Montagnac, and P.~Sagaut.
\newblock {Towards an adaptive POD/SVD surrogate model for aeronautic design}.
\newblock \emph{Computers {\&} Fluids}, 40\penalty0 (1):\penalty0 195--209,
  2011.
\newblock \doi{10.1016/j.compfluid.2010.09.002}.

\bibitem[{Chatterjee, Anindya}(2000)]{chatterjee2000}
{Chatterjee, Anindya}.
\newblock {An introduction to the proper orthogonal decomposition}.
\newblock \emph{Current Science}, 78\penalty0 (7), 2000.

\bibitem[Clarke et~al.(1992)Clarke, Fokoue, and Zhang]{clarke1992}
Bertrand Clarke, Ernest Fokoue, and Hao~Helen Zhang.
\newblock \emph{{Breakthroughs in Statistics}}.
\newblock Springer Series in Statistics. Springer New York, 1992.
\newblock ISBN 978-0-387-94039-7.
\newblock \doi{10.1007/978-1-4612-4380-9}.

\bibitem[Cloke and Pappenberger(2009)]{cloke2009}
H.L. Cloke and F.~Pappenberger.
\newblock {Ensemble flood forecasting: A review}.
\newblock \emph{Journal of Hydrology}, 375:\penalty0 613--626, 2009.
\newblock ISSN 00221694.
\newblock \doi{10.1016/j.jhydrol.2009.06.005}.

\bibitem[Damblin et~al.(2013)Damblin, Couplet, Iooss, Damblin, Couplet, and
  Iooss]{damblin2013}
Guillaume Damblin, Mathieu Couplet, Bertrand Iooss, Guillaume Damblin, Mathieu
  Couplet, and Bertrand Iooss.
\newblock {Numerical studies of space filling designs : optimization of Latin
  Hypercube Samples and subprojection properties}.
\newblock \emph{Journal of Simulation}, pages 276--289, 2013.

\bibitem[Dechant and Moradkhani(2011)]{dechant2011}
C.M. Dechant and H.~Moradkhani.
\newblock {Improving the characterization of initial condition for ensemble
  streamflow prediction using data assimilation}.
\newblock \emph{Hydrology and Earth System Sciences}, 15:\penalty0 3399--3410,
  2011.
\newblock \doi{10.5194/hess-15-3399-2011}.

\bibitem[Deman et~al.(2015)Deman, Konakli, Sudret, Kerrou, Perrochet, and
  Benabderrahmane]{deman2015}
G.~Deman, K.~Konakli, B.~Sudret, J.~Kerrou, P.~Perrochet, and
  H.~Benabderrahmane.
\newblock Using sparse polynomial chaos expansions for the global sensitivity
  analysis of groundwater lifetime expectancy in a multi-layered
  hydrogeological model.
\newblock \emph{Reliability Engineering and System Safety}, 147:\penalty0
  156--169, 2015.

\bibitem[Despr{\'e}s et~al.(2013)Despr{\'e}s, Poette, and Lucor]{despres2013}
Bruno Despr{\'e}s, Gael Poette, and Didier Lucor.
\newblock \emph{Robust Uncertainty Propagation in Systems of Conservation Laws
  with the Entropy Closure Method}, pages 105--149.
\newblock Springer International Publishing, Cham, 2013.
\newblock \doi{10.1007/978-3-319-00885-1\_3}.

\bibitem[Dubreuil et~al.(2014)Dubreuil, Berveiller, Petitjean, and
  Salaün]{dubreuil2014}
S.~Dubreuil, M.~Berveiller, F.~Petitjean, and M.~Salaün.
\newblock Construction of bootstrap confidence intervals on sensitivity indices
  computed by polynomial chaos expansion.
\newblock \emph{Reliability Engineering and System Safety}, 121:\penalty0
  263--275, 2014.
\newblock \doi{10.1016/j.ress.2013.09.011}.

\bibitem[Durand et~al.(2008)Durand, Andreadis, Alsdorf, Lettenmaier, Moller,
  and Wilson]{durand2008}
M.~Durand, K.M. Andreadis, D.E. Alsdorf, D.P. Lettenmaier, D.~Moller, and
  M.~Wilson.
\newblock {Estimation of bathymetric depth and slope from data assimilation of
  swath altimetry into a hydrodynamic model}.
\newblock \emph{Geophysical Research Letters}, 35:\penalty0 1--5, 2008.
\newblock \doi{10.1029/2008GL034150}.

\bibitem[Dutka-Malen et~al.(2009)Dutka-Malen, Lebrun, Saassouh, and
  Sudret]{dutka2009}
I.~Dutka-Malen, R.~Lebrun, B.~Saassouh, and B.~Sudret.
\newblock Implementation of a polynomial chaos toolbox in openturns with
  test-case application.
\newblock In \emph{Conference: Proc. 10th~Int. Conf. Struct. Safety and
  Reliability (ICOSSAR'2009), Osaka, Japan}, 2009.

\bibitem[El~Mo\c{c}ayd et~al.()El~Mo\c{c}ayd, Ricci, Goutal, Rochoux, Boyaval,
  Goeury, Lucor, and Thual]{elmocaydEMA}
N.~El~Mo\c{c}ayd, S.~Ricci, N.~Goutal, M.C. Rochoux, S.~Boyaval, C.~Goeury,
  D.~Lucor, and O.~Thual.
\newblock {Polynomial surrogate model for open-channel flows in steady state}.

\bibitem[Goutal and Maurel(2002)]{goutal2002}
N.~Goutal and F.~Maurel.
\newblock {A finite volume solver for 1D shallow-water equations applied to an
  actual river}.
\newblock \emph{International Journal for Numerical Methods in Fluids},
  38\penalty0 (1):\penalty0 1--19, 2002.

\bibitem[Goutal et~al.(2012)Goutal, Lacombe, Zaoui, and K.]{goutal2012}
N.~Goutal, J.-M. Lacombe, F.~Zaoui, and El-Kadi-Adberrezzak K.
\newblock Mascaret: a 1-d open souces software for flow hydrodynamic and water
  quality in open channel networks.
\newblock \emph{River Flow}, pages 1169--1174, 2012.

\bibitem[Habert et~al.(2016)Habert, Ricci, {Le Pape}, Thual, Piacentini,
  Goutal, Jonville, and Rochoux]{habert2016}
J.~Habert, S.~Ricci, E.~{Le Pape}, O.~Thual, A.~Piacentini, N.~Goutal,
  G.~Jonville, and M.~Rochoux.
\newblock {Reduction of the uncertainties in the water level-discharge relation
  of a 1D hydraulic model in the context of operational flood forecasting}.
\newblock \emph{Journal of Hydrology}, 532:\penalty0 52--64, 2016.
\newblock \doi{10.1016/j.jhydrol.2015.11.023}.

\bibitem[Hastie et~al.(2009)Hastie, Tibshirani, and Friedman]{hastie2009}
Trevor Hastie, Robert Tibshirani, and Jerome Friedman.
\newblock \emph{{The Elements of Statistical Learning}}, volume~2 of
  \emph{Springer Series in Statistics}.
\newblock Springer New York, New York, NY, 2009.
\newblock ISBN 978-0-387-84857-0.
\newblock \doi{10.1007/978-0-387-84858-7}.

\bibitem[Horritt and Bates(2002)]{horritt2002}
M.S. Horritt and P.D. Bates.
\newblock Evaluation of 1d and 2d numerical models for predicting river flood
  inundation.
\newblock \emph{Journal of Hydrology}, 268:\penalty0 87--99, 2002.

\bibitem[Hosder and Walters(2006)]{hosder2006}
R.~Hosder, S.~Perez and R.W. Walters.
\newblock A non-intrusive polynomial chaos method for uncertainty propagation
  in cfd simulations.
\newblock In \emph{48th AIAA Aerospace Sciences Meeting and Exhibit}, number
  AIAA-2010-0129. The American Institute of Aeronautics and Astronautics, Inc,
  2006.

\bibitem[Iooss and Saltelli(2016)]{iooss2016}
B.~Iooss and A.~Saltelli.
\newblock {Introduction to Sensitivity Analysis}.
\newblock In \emph{Handbook of Uncertainty Quantification}, pages 1--20.
  Springer International Publishing, 2016.
\newblock \doi{10.1007/978-3-319-11259-6\_31-1}.

\bibitem[Iooss et~al.(2010)Iooss, Boussouf, Feuillard, and Marrel]{iooss2010}
B.~Iooss, L.~Boussouf, V.~Feuillard, and A.~Marrel.
\newblock {Numerical studies of the metamodel fitting and validation
  processes}.
\newblock \emph{International Journal on Advances in Systems and Measurements},
  3\penalty0 (1):\penalty0 11--21, 2010.

\bibitem[Ishigami and Homma(1990)]{ishigami1990}
T.~Ishigami and T.~Homma.
\newblock {An importance quantification technique in uncertainty analysis for
  computer models}.
\newblock \emph{IEEE}, pages 398--403, 1990.
\newblock \doi{10.1109/ISUMA.1990.151285}.

\bibitem[Lamboni et~al.(2011)Lamboni, Monod, and Makowski]{lamboni2011}
M.~Lamboni, H.~Monod, and D.~Makowski.
\newblock {Multivariate sensitivity analysis to measure global contribution of
  input factors in dynamic models}.
\newblock \emph{Reliability Engineering and System Safety}, 96\penalty0
  (4):\penalty0 450--459, 2011.
\newblock \doi{10.1016/j.ress.2010.12.002}.

\bibitem[Le~Gratiet et~al.(2014)Le~Gratiet, Cannamela, and
  Iooss]{legratiet2014}
L.~Le~Gratiet, C.~Cannamela, and B.~Iooss.
\newblock A bayesian approach for global sensitivity analysis of
  (multifidelity) computer codes.
\newblock \emph{SIAM/ASA Journal on Uncertainty Quantification}, 2\penalty0
  (1):\penalty0 336--363, 2014.
\newblock \doi{10.1137/130926869}.

\bibitem[{Le Gratiet} et~al.(2017){Le Gratiet}, Marelli, and
  Sudret]{legratiet2017}
L.~{Le Gratiet}, S.~Marelli, and B.~Sudret.
\newblock {Metamodel-Based Sensitivity Analysis: Polynomial Chaos Expansions
  and Gaussian Processes}.
\newblock In \emph{Handbook of Uncertainty Quantification}, pages 1--37.
  Springer International Publishing, 2017.
\newblock \doi{10.1007/978-3-319-11259-6\_38-1}.

\bibitem[Le~Maitre and Knio(2010)]{lemaitreknio2010}
O.~Le~Maitre and O.~Knio.
\newblock \emph{Spectral Methods for Uncertainty Quantification}.
\newblock Springer, 2010.

\bibitem[Li and Xiu(2008)]{lixiu2008}
J.~Li and D.~Xiu.
\newblock On numerical properties of the ensemble {K}alman filter for data
  assimilation.
\newblock \emph{Comput. Methods Appl. Mech. Engrg.}, 197:\penalty0 3574--3583,
  2008.

\bibitem[Li and Xiu(2009)]{lixiu2009}
J.~Li and D.~Xiu.
\newblock A generalized polynomial chaos based ensemble kalman filter with high
  accuracy.
\newblock \emph{Journal of Computational Physics}, 228\penalty0 (15):\penalty0
  5454--5469, 2009.
\newblock \doi{10.1016/j.jcp.2009.04.029}.

\bibitem[Liang et~al.(2008)Liang, Kwok~Fai, and Kobayashi]{ge2008}
Ge~Liang, Cheung Kwok~Fai, and Marcelo~H. Kobayashi.
\newblock Stochastic solution for uncertainty propagation in nonlinear
  shallow-water equations.
\newblock \emph{Journal of Hydraulic Engineering}, 134\penalty0 (12):\penalty0
  1732--1743, 2008.
\newblock \doi{10.1061/(ASCE)0733-9429(2008)134:12(1732)}.

\bibitem[Lockwood and Anitescu(2012)]{lockwood2012}
B.A. Lockwood and M.~Anitescu.
\newblock {Gradient-enhanced universal kriging for uncertainty propagation}.
\newblock \emph{Nuclear Science and Engineering}, pages 1--32, 2012.

\bibitem[Lucor et~al.(2007)Lucor, Meyers, and Sagaut]{lucor2007}
D.~Lucor, J.~Meyers, and P.~Sagaut.
\newblock {Sensitivity analysis of large-eddy simulations to
  subgrid-scale-model parametric uncertainty using polynomial chaos}.
\newblock \emph{Journal of Fluid Mechanics}, 585:\penalty0 255--279, 2007.
\newblock \doi{10.1017/S0022112007006751}.

\bibitem[Marrel et~al.(2009)Marrel, Iooss, Laurent, and Roustant]{marrel2009}
Amandine Marrel, Bertrand Iooss, Beatrice Laurent, and Olivier Roustant.
\newblock Calculations of sobol indices for the gaussian process metamodel.
\newblock \emph{Reliability Engineering \& System Safety}, 94\penalty0
  (3):\penalty0 742 -- 751, 2009.
\newblock \doi{http://dx.doi.org/10.1016/j.ress.2008.07.008}.

\bibitem[Matgen et~al.(2010)Matgen, Montanari, Hostache, Pfister, Hoffmann,
  Plaza, Pauwels, {De Lannoy}, {De Keyser}, and Savenije]{matgen2010}
P.~Matgen, M.~Montanari, R.~Hostache, L.~Pfister, L.~Hoffmann, D.~Plaza, V.R.N.
  Pauwels, G.J.M. {De Lannoy}, R.~{De Keyser}, and H.H.G. Savenije.
\newblock {Towards the sequential assimilation of SAR-derived water stages into
  hydraulic models using the Particle Filter: Proof of concept}.
\newblock \emph{Hydrology and Earth System Sciences}, 14:\penalty0 1773--1785,
  2010.
\newblock \doi{10.5194/hess-14-1773-2010}.

\bibitem[Migliorati et~al.(2013)Migliorati, Nobile, Von~Schwerin, and
  Tempone]{migliorati2013}
G.~Migliorati, F.~Nobile, E.~Von~Schwerin, and R.~Tempone.
\newblock {Approximation of quantities of interest in stochastic PDEs by the
  random Discret L2 Projection on polynomial spaces}.
\newblock \emph{SIAM J. Sci Comput.}, 35\penalty0 (3):\penalty0 A1440--A1460,
  2013.

\bibitem[Molga and Smutnicki(2005)]{molga2005}
Marcin Molga and Czes{\l}aw Smutnicki.
\newblock {Test functions for optimization needs}.
\newblock \penalty0 (c):\penalty0 1--43, 2005.

\bibitem[Moradkhani et~al.(2005)Moradkhani, Sorooshian, Gupta, and
  Houser]{moradkhani2005}
H.~Moradkhani, S.~Sorooshian, H.V. Gupta, and P.R. Houser.
\newblock {Dual state-parameter estimation of hydrological models using
  ensemble Kalman filter}.
\newblock \emph{Advances in Water Resources}, 28:\penalty0 135--147, 2005.
\newblock \doi{10.1016/j.advwatres.2004.09.002}.

\bibitem[Oakley and O'Hagan(2004)]{oakley2004}
J.E. Oakley and A.~O'Hagan.
\newblock Probabilistic sensitivity analysis of complex models: a bayesian
  approach.
\newblock \emph{Journal of the Royal Statistical Society: Series B (Statistical
  Methodology)}, 66\penalty0 (3):\penalty0 751--769, 2004.
\newblock \doi{10.1111/j.1467-9868.2004.05304.x}.

\bibitem[Owen et~al.(2015)Owen, Challenor, Menon, and Bennani]{owen2015}
N.E. Owen, P.~Challenor, P.~P. Menon, and S.~Bennani.
\newblock {Comparison of surrogate-based uncertainty quantification methods for
  computationally expensive simulators}.
\newblock pages 1--10, 2015.
\newblock URL \url{http://arxiv.org/abs/1511.0926
  http://arxiv.org/abs/1511.00926}.

\bibitem[Parrish et~al.(2012)Parrish, Moradkhani, and Dechant]{parrish2012}
M.A. Parrish, H.~Moradkhani, and C.M. Dechant.
\newblock {Toward reduction of model uncertainty: Integration of Bayesian model
  averaging and data assimilation}.
\newblock \emph{Water Resources Research}, 48\penalty0 (W03519):\penalty0
  1--18, 2012.
\newblock \doi{10.1029/2011WR011116}.

\bibitem[Pedregosa et~al.(2012)Pedregosa, Varoquaux, Gramfort, Michel, Thirion,
  Grisel, Blondel, Prettenhofer, Weiss, Dubourg, Vanderplas, Passos,
  Cournapeau, Brucher, Perrot, and Duchesnay]{pedregosa2011}
Fabian Pedregosa, Ga{\"{e}}l Varoquaux, Alexandre Gramfort, Vincent Michel,
  Bertrand Thirion, Olivier Grisel, Mathieu Blondel, Peter Prettenhofer, Ron
  Weiss, Vincent Dubourg, Jake Vanderplas, Alexandre Passos, David Cournapeau,
  Matthieu Brucher, Matthieu Perrot, and {\'{E}}douard Duchesnay.
\newblock {Scikit-learn: Machine Learning in Python}.
\newblock \emph{Journal of Machine Learning Research}, 12\penalty0 (2825-2830),
  2012.

\bibitem[Rasmussen and Williams(2006)]{rasmussen2006}
C.E. Rasmussen and C~Williams.
\newblock \emph{{Gaussian processes for machine learning}}.
\newblock MIT Press, 2006.

\bibitem[Rochoux et~al.(2014)Rochoux, Ricci, Lucor, and Trouv\'e]{rochoux2014}
M.C. Rochoux, S.~Ricci, B.~Lucor, D.and~Cuenot, and A.~Trouv\'e.
\newblock {Towards predictive data-driven simulations of wildfire spread - Part
  1: Reduced-cost Ensemble Kalman Filter based on a Polynomial Chaos surrogate
  model for parameter estimation}.
\newblock \emph{Nat. Hazards and Earth Syst. Sci.}, 14\penalty0 (11):\penalty0
  2951--2973, 2014.

\bibitem[Roy(2016)]{roy2016}
Pamphile~T. Roy.
\newblock {Uncertainty Quantification applied to Turbine Design}.
\newblock Technical report, CERFACS, Toulouse, 2016.

\bibitem[Saad(2007)]{saad2007phd}
G.A. Saad.
\newblock \emph{{Stochastic Data Assimilation with Application to Multi-Phase
  Flow and Health Monitoring Problems}}.
\newblock PhD thesis, Faculty of the Graduate School, University of Southern
  California, 2007.

\bibitem[Saltelli et~al.(2007)Saltelli, Ratto, Andres, Campolongo, Cariboni,
  Gatelli, Saisana, and Tarantola]{saltelli2007}
Andrea Saltelli, Marco Ratto, Terry Andres, Francesca Campolongo, Jessica
  Cariboni, Debora Gatelli, Michaela Saisana, and Stefano Tarantola.
\newblock \emph{{Global Sensitivity Analysis. The Primer}}.
\newblock John Wiley {\&} Sons, Ltd, Chichester, UK, dec 2007.
\newblock \doi{10.1002/9780470725184}.

\bibitem[Schoebi et~al.(2015)Schoebi, Sudret, and Wiart]{schoebi2015}
R.~Schoebi, B.~Sudret, and J.~Wiart.
\newblock {Polynomial-Chaos-based Kriging}.
\newblock \emph{International Journal for Uncertainty Quantification},
  5\penalty0 (2):\penalty0 171--193, 2015.

\bibitem[Sirovich(1987)]{sirovich1987}
Lawrence Sirovich.
\newblock {Turbulence and the dynamics of coherent structures part i: coherent
  structures}.
\newblock \emph{Quarterly of Applied Mathematics}, XLV\penalty0 (3):\penalty0
  561--571, 1987.

\bibitem[Smirnov(1939)]{smirnov1939}
NV~Smirnov.
\newblock Estimate of difference between empirical distribution curves in two
  independent samples.
\newblock \emph{Byull. Mosk. Gos. Univ}, 2\penalty0 (2), 1939.

\bibitem[Sobol′(1993)]{sobol1993}
I.M Sobol′.
\newblock {Sensitivity analysis for nonlinear mathematical models}.
\newblock \emph{Mathematical Modeling and Computational Experiment}, 1\penalty0
  (4):\penalty0 407--414, 1993.

\bibitem[Storlie et~al.(2009)Storlie, Swiler, Helton, and
  Sallaberry]{storlie2009}
C.B. Storlie, L.P. Swiler, J.C. Helton, and C.J. Sallaberry.
\newblock {Implementation and evaluation of nonparametric regression procedures
  for sensitivity analysis of computationally demanding models}.
\newblock \emph{Reliability Engineering {\&} System Safety}, 94\penalty0
  (11):\penalty0 1735--1763, nov 2009.
\newblock \doi{10.1016/j.ress.2009.05.007}.

\bibitem[Sudret(2008)]{sudret2008}
B.~Sudret.
\newblock Global sensitivity analysis using polynomial chaos expansions.
\newblock \emph{Reliability Engineering and System Safety}, 93\penalty0
  (7):\penalty0 964--979, 2008.
\newblock \doi{10.1016/j.ress.2007.04.002}.

\bibitem[Thual(2010)]{thual2010}
O.~Thual.
\newblock \emph{{Hydrodynamique de l'environnement}}.
\newblock Ecole polytechnique, 2010.

\bibitem[Wales and Doye(1997)]{wales1997}
David~J. Wales and Jonathan P.~K. Doye.
\newblock {Global Optimization by Basin-Hopping and the Lowest Energy
  Structures of Lennard-Jones Clusters Containing up to 110 Atoms}.
\newblock \emph{The Journal of Physical Chemistry A}, 101\penalty0
  (28):\penalty0 5111--5116, 1997.
\newblock \doi{10.1021/jp970984n}.

\bibitem[Wand and Jones(1995)]{wand1995}
M.~P. Wand and M.~C. Jones.
\newblock \emph{{Kernel Smoothing}}.
\newblock Springer US, Boston, MA, 1995.
\newblock ISBN 978-0-412-55270-0.
\newblock \doi{10.1007/978-1-4899-4493-1}.

\bibitem[Weerts et~al.(2011)Weerts, Winsemius, and Verkade]{weerts2011}
A.H. Weerts, H.C. Winsemius, and J.S. Verkade.
\newblock {Estimation of predictive hydrological uncertainty using quantile
  regression: examples from the National Flood Forecasting System (England and
  Wales)}.
\newblock \emph{Hydrology and Earth System Sciences}, 15:\penalty0 255--265,
  2011.
\newblock \doi{10.5194/hess-15-255-2011}.

\bibitem[Xiu(2010)]{xiu2010}
D.~Xiu.
\newblock \emph{Numerical Methods for Stochastic Computations: A Spectral
  Method Approach}.
\newblock Princeton University Press, 2010.

\bibitem[Xiu and Karniadakis(2002)]{xiu2002}
D.~Xiu and G.E. Karniadakis.
\newblock The wiener--askey polynomial chaos for stochastic differential
  equations.
\newblock \emph{SIAM Journal on Scientific Computing}, 24\penalty0
  (2):\penalty0 619--644, 2002.
\newblock \doi{10.1137/S1064827501387826}.

\end{thebibliography}

\end{document}